\documentclass[%
 aip,
 jmp,%
 amsmath,amssymb,
reprint,%
]{revtex4-1}

\usepackage{graphicx}
\usepackage{dcolumn}
\usepackage{bm}

\def\lapprox{\mathrel{\hbox{\rlap{\hbox{\lower4pt\hbox{$\sim$}}}\hbox{$<$}}}}
\def\gapprox{\mathrel{\hbox{\rlap{\hbox{\lower4pt\hbox{$\sim$}}}\hbox{$>$}}}}

\newcommand{\be}{\begin{equation}}
\newcommand{\ee}{\end{equation}}

\newcommand {\nind} {\noindent}
\newcommand {\mb} {\mathbf}

\newcommand {\bea} {\begin{eqnarray}}
\newcommand {\eea} {\end{eqnarray}}

\begin{document}

\title{Magnetic Reconnection in a Weakly Ionized Plasma}

\author{James E. Leake}
\email{jleake@gmu.edu}
\affiliation{College of Science, George Mason University, 4400 University Drive, Fairfax, Virginia 22030.}
\author{Vyacheslav S. Lukin, Mark G. Linton}%
\affiliation{ U.S. Naval Research Lab 4555 Overlook Ave., SW Washington, DC 20375}

\date{\today}

\begin{abstract}
Magnetic reconnection in partially ionized plasmas is a ubiquitous phenomenon spanning the range from laboratory to intergalactic scales, yet it remains poorly understood and relatively little studied.  Here, we present results from a self-consistent multi-fluid simulation of magnetic reconnection in a weakly ionized reacting plasma with a particular focus on the parameter regime of the solar chromosphere. The numerical model includes collisional transport, interaction and reactions between the species, and optically thin radiative losses. This model improves upon our previous work in Leake et al. 2012 \cite{Leake2012} by considering realistic chromospheric transport coefficients, and by solving a generalized Ohm's law that accounts for finite ion-inertia and electron-neutral drag.
 We find that during the two dimensional reconnection of a Harris current sheet with an initial width larger than the neutral-ion collisional coupling scale, the current sheet thins until its width becomes less than this coupling scale, and the neutral and ion fluids decouple upstream from the reconnection site.  During this process of decoupling, we observe reconnection faster than the single-fluid Sweet-Parker prediction, with recombination and plasma outflow both playing a role in determining the reconnection rate. 
  As the current sheet thins further and elongates it becomes unstable to the secondary tearing instability, and plasmoids are seen. The reconnection rate, outflows and plasmoids observed in this simulation provide evidence that magnetic reconnection in the chromosphere could be responsible for jet-like transient phenomena such as spicules and chromospheric jets.
\end{abstract}

\maketitle
\section{Introduction}

Magnetic reconnection, the process which converts energy stored in the magnetic field into kinetic and thermal energy of the plasma by breaking magnetic connectivity, is observed in a range of laboratory and astrophysical plasmas. In the astrophysical context, these include the interstellar medium (ISM), galactic disks, and the heliosphere \cite{2009ARA&A..47..291Z}.  In the heliosphere in particular, magnetic reconnection is believed to be the cause of transient phenomena such as solar flares and X-ray jets \cite{2011ApJ...731L..18M}, magnetospheric substorms \cite{2007JGRA..11206215B,2007JGRA..11201209M} and solar $\gamma$-ray bursts \cite{2005JGRA..11011103E,2005ApJ...628L..77T}. 

Reconnection occurs when the ideal MHD frozen-in constraint, valid for highly conducting plasmas, is broken locally, by allowing fieldlines to reconnect through a narrow diffusion region.
Parker and Sweet (Refs.~\onlinecite{1957JGR....62..509P, Sweet58}) were the first to formulate magnetic reconnection as a local process within the single-fluid, fully ionized, magnetohydrodynamic (MHD) framework. They considered a current layer of width $\delta_{SP}$ much smaller than its length $L$, and plasma with non-zero resistivity ($\eta$) due to electron-ion collisions. The model assumes steady state, i.e.,  the 
rate of supply of ions to the reconnection region by inflow is equal to the rate of removal of ions by outflow, and takes the length $L$ to be the characteristic system length scale.  Using Ohm's law for a fully ionized, single fluid plasma with velocity $\mb{v}$, magnetic field $\mb{B}$, electric field $\mb{E}$, and current density $\mb{j}$:
\be
\mathbf{E}+\mathbf{v}\times\mathbf{B}  = \eta \mathbf{j},
\ee
along with the plasma momentum equation and the steady state continuity equation, a simple equation for the reconnection rate can be derived:
\be
M \equiv \frac{v_{in}}{v_{A}} \approx \sqrt{\frac{\eta}{\mu_{0}v_{A}L}} = \frac{1}{\sqrt{S}}.
\ee
Here $S=\mu_{0}v_{A}L/\eta$ is the Lundquist number, $v_{in}$ is the ion inflow upstream from the current sheet,  $v_{A}=B/\sqrt{\rho\mu_{0}}$ is a typical Alfv\'{e}n velocity, with $B$  evaluated upstream from the current sheet and $\rho$ (mass density) evaluated in the current sheet. The permeability of free space is denoted by $\mu_{0}$. From this analysis, it can also be shown that the Sweet-Parker current width $\delta_{SP}$ and aspect ratio $\sigma_{SP} \equiv \delta_{SP}/L$ approximate to
\be
\delta_{SP} \approx \sqrt{\frac{\eta L}{\mu_{0}v_{A}}}, ~ \sigma_{SP} \approx \frac{1}{\sqrt{S}}.
\label{eqn:width}
\ee

In this paper, we will focus on magnetic reconnection in the solar chromosphere, which exhibits localized, transient outflows on a number of different length-scales. ``Chromospheric jets''  are the largest class of these outflows, and are 
surges of plasma with typical lifetimes of 200-1000 s, lengths of 5 Mm, and velocities at their base of 10 km/s (e.g., Refs. \onlinecite{2007Sci...318.1591S, 2012ApJ...760...28S,2012ApJ...759...33S}). ``Spicules" are a smaller class of chromospheric outflows and 
have lifetimes of 10-600 s, lengths of up to 1 Mm and velocities of 20-150 km/s (e.g., Ref. \onlinecite{2000SoPh..196...79S}). A unified model for the generation 
of chromospheric outflows was suggested by Refs. \onlinecite{2007Sci...318.1591S,2011ApJ...731L..18M}. In their model, a bipolar field emerging into and reconnecting with a unipolar field creates outwardly directed reconnection jets.  Depending on the size of the bipole and the strength of the field, this model may be able to explain some of the observed properties of spicules and chromospheric jets. In Leake et al. 2012 \cite{Leake2012}  we argued that, given the observed lifetimes, lengths and velocities of chromospheric jets, the minimum normalized reconnection rate is $M\approx 0.5$. For spicules the minimum value we derived is $M\approx 0.01$.

The solar chromosphere is a weakly ionized plasma, where the average collision time between neutrals and ions is on the order of ms, and so these two fluids can generally be considered to act as a single fluid. In this regime, the Sweet-Parker reconnection scaling is applicable, though the Alfv\'{e}n speed, $v_{A}$, now depends on the total (ion+neutral) mass. 
However, if the width of the current sheet is comparable to the neutral-ion collisional mean free path, $\lambda_{ni}\equiv v_{T,n}/\nu_{ni}$ (where $v_{T,n}$ is the thermal velocity of neutrals and $\nu_{ni}$ is the neutral-ion collisional frequency), ions and neutrals can decouple and the mix
may not be in ionization balance. In this case the reconnection region will have sources and sinks of ions in the form of the atomic reactions of ionization and recombination, and the inflow and outflow rates for the reconnection region could be affected by these interactions.

The model for the multi-fluid simulations we presented in Leake et al. 2012 \cite{Leake2012} included ionization, recombination and scattering collisions, but did not include charge exchange collisions. We
found fast reconnection occurred when the ions and neutrals became decoupled upstream from the reconnection site, and the resulting recombination of excess ions within the current sheet was as efficient at removing ions as the expulsion by reconnecting field. In this paper we improve upon that model in a number of ways. Firstly, we include charge exchange collisions and use more realistic cross-sections for scattering collisions, which effectively increases the collisional coupling of neutrals and ions, and thus decreases the neutral-ion collisional mean free path. At the same time, we also have a higher Lundquist number (see below) than in Leake et al. 2012 \cite{Leake2012} which reduces the Sweet-Parker width. As we will show, these changes result in similar ratios of the Sweet-Parker width to the neutral-ion collisional scale. We suggest that it is this ratio that  is important for obtaining fast reconnection, and so the simulation in this paper and those in Leake et al. 2012 \cite{Leake2012} are consistent despite these changes. Secondly, we allow the neutral fluid collisional transport coefficients to be functions of the local plasma conditions, rather than model parameters, and use more realistic values for the fluid transport coefficients. This results in a magnetic Prandtl number, the ratio of viscous to resistive diffusivity, which is lower than the simulations of Leake et al. 2012 \cite{Leake2012}.

Thirdly, we solve a generalized Ohm's law that accounts for finite ion inertia and electron-neutral drag effects. In the simulations of Leake et al. 2012 \cite{Leake2012} the resistivity was a parameter of the simulations, and we derived scaling relationships for the reconnection rate and current sheet width in terms of the resistivity (or Lundquist number). The resistivity used in this paper, which appropriately depends on
 the plasma density and temperature, is a more realistic estimate for the chromospheric resistivity, and hence Lundquist number. The numerical model used in this paper is presented in \S2 and the results are presented in \S 3. \S 4 provides discussion of the results in reference to the transient phenomena observed in the chromosphere such as spicules and jets.

\section{Numerical Method}

\subsection{Multi-Fluid Partially Ionized Plasma Model}
\label{sec:HiFi_PN}
The chromospheric model is based on the model used in our previous simulations in Leake et al. 2012 \cite{Leake2012}, and so we only present here the modifications made to that model, and the necessary information to understand the modifications. The model consists of three fluids, ion (\textit{i}),
electron (\textit{e}), and neutral (\textit{n}). Each fluid ($\alpha$) has a number density $n_{\alpha}$, particle mass $m_{\alpha}$, pressure tensor $\mathbb{P}_{\alpha}$, velocity $\mb{v}_{\alpha}$, and temperature $T_{\alpha}$. Only hydrogen is considered here, and so $m_{i}$ and $m_{n}$ are equal to the mass of a proton. These fluids can undergo  
recombination, ionization and charge exchange interactions.  The recombination and  ionization reaction rates are given by Equations (4)-(7) in Leake et al. 2012 \cite{Leake2012}. As mentioned, in Leake et al. 2012 \cite{Leake2012} we did not include charge exchange interactions, but do so in the simulation presented in this paper.

The charge exchange reaction rate $\Gamma^{cx}$ is defined as 
\be
\Gamma^{cx} \equiv \Sigma_{cx}(V_{cx})n_{i}n_{n}V_{cx}, \ee 
where \be
V_{cx}\equiv
\sqrt{\frac{4}{\pi}v_{Ti}^{2}+\frac{4}{\pi}v_{Tn}^{2}+v_{in}^{2}} \ee
is the representative speed of the interaction and $v_{in}^{2} \equiv
{|\mb{v}_{i}-\mb{v}_{n}|}^2$. The thermal speed of species $\alpha$ is given by
$v_{T\alpha} = \sqrt{2k_{B}T_{\alpha}/m_{\alpha}}$, where $k_{B}$ is Boltzmann's constant.  Following 
Ref. \onlinecite{Meier11}, the charge exchange cross-section $\Sigma_{cx}(V_{cx})$ is given by
\be
\Sigma_{cx}(V_{cx}) = 1.12\times10^{-18} - 7.15\times10^{-20}\ln{V_{cx}} ~ \textrm{m}^{2}
\ee
which is a practical fit to observational data of Hydrogen collisions \citep{1990STIN...9113238B}.

\nind\textit{Continuity: } \\
The continuity equations for the ions and neutrals are given by Equations (10) and (11) in Leake et al. 2012 \cite{Leake2012}, respectively, and include ionization and radiative recombination. The electron continuity equation is not required as charge neutrality ($n_{e}=n_{i}$) is assumed.

\nind\textit{Momentum: } \\
The ionized fluid (electron+ion), and neutral fluid momentum equations are given by Equation (12) and (13) in Leake et al. 2012 \cite{Leake2012}, respectively, and include ionization, recombination and charge exchange contributions. As in Leake et al. 2012 \cite{Leake2012}, we neglect the viscous part of the electron pressure tensor, but in this paper the contributions due to charge exchange are retained.

In addition, we use different collisional cross-sections for ions and neutrals, compared to the model in Leake et al. 2012 \cite{Leake2012}. The collision frequency for collisions of ions with neutrals is 
$\nu_{i n}$ is given by
\be
\nu_{i n} = n_{n}\Sigma_{i n}\sqrt{\frac{8k_{B}T_{i n}}{\pi m_{i n}}}
\ee
with $T_{\alpha\beta} = \frac{T_{\alpha}+T_{\beta}}{2}$ and $m_{\alpha\beta}  = \frac{m_{\alpha}m_{\beta}}{m_{\alpha}+m_{\beta}}$. The ion neutral collisional cross-section is $\Sigma_{i n}$. In this paper we use $\Sigma_{in} = \Sigma_{ni} = 1.16\times 10^{-18} ~ \textrm{m}^2$, whereas in Leake et al. 2012 \cite{Leake2012}, the model used  $\Sigma_{in} = \Sigma_{ni} = 1.4\times 10^{-19} ~ \textrm{m}^2$. These improvements are made to take into account the theoretical work of Ref. \onlinecite{1983ApJ...264..485D}.

In the model in this paper we also use plasma dependent viscosity coefficients in the pressure tensor, unlike in Leake et al. 2012 \cite{Leake2012} where they were merely parameters (the neutral viscosity coefficient $\xi_{n}$ was set to $10^{-3} ~ \textrm{kg}/(\textrm{m}.\textrm{s})$, as was the ion viscosity coefficient). In this paper
the neutral and ion viscosity coefficients are given by
\be
\xi_{n} = \frac{n_{n}k_{B}T_{n}}{\nu_{nn}}, ~ \textrm{and} ~  \xi_{i} = \frac{n_{i,0}k_{B}T_{i,0}}{ {\nu_{ii}}_0},
\ee
respectively. The relevant collision frequencies $\nu_{nn}$ and $\nu_{ii}$ are given by
\be
\nu_{n n} = n_{n}\Sigma_{n n}\sqrt{\frac{16k_{B}T_{n}}{\pi m_{n}}}, ~ \textrm{and} ~
\nu_{i i} = \frac{4}{3}n_{i}\Sigma_{i i}\sqrt{\frac{2k_{B}T_{i}}{\pi m_{i}}}
\ee
where $\Sigma_{nn}=7.73\times10^{-19} ~ \textrm{m}^{2}$, and $\Sigma_{i i} = \lambda \pi r_{d,i}^{2}$, with $r_{d,i}=e^{2}/(4\pi\epsilon_{0} k_{B}T_{i})$ being the distance of closest approach for ions ($\epsilon_{0}$ is the permittivity of free space, $e$ is the elementary charge, and $\lambda=10$ is the Coulomb logarithm). The subscripts $0$ in the ion coefficient represent background values at time $t=0$. 
Hence the neutral viscosity coefficient varies but the ion viscosity coefficient has a constant value equivalent to unmagnetized ion viscosity in a uniform background plasma of temperature $T_{i,0}$ and density $n_{i,0}$. Future work will consider the effects of an anisotropic ion stress tensor.

\nind\textit{Internal Energy:} \\
The internal energy equations for the ionized fluid (ion+electron) and the neutral fluid are given by Equations (19) and (20) in Leake et al. 2012 \cite{Leake2012}, respectively, and include ionization, recombination, charge exchange, and optically thin radiative losses.  In this paper we retain the charge-exchange interaction contributions. We also use a neutral thermal conductivity ($\kappa_{n}$) that is dependent on the plasma parameters 
\be
\kappa_{n} = \frac{4n_{n}k_{B}T_{n}}{\nu_{nn}m_{i}}.
\ee
This has a value of $\kappa_{n}=1.76\times10^{24} ~ \textrm{m}^{-1}\textrm{s}^{-1}$ for the initial background neutral density and temperature used in the simulation in this paper. In Leake et al. 2012 \cite{Leake2012} we used a constant value, independent of the neutral fluid conditions, of $\kappa_{n}=1.58\times10^{24} ~ \textrm{m}^{-1}\textrm{s}^{-1}$.
Note that the frictional heating and thermal transfer term $Q_{\alpha}^{\alpha\beta}$ was quoted incorrectly in Leake et al. 2012 \cite{Leake2012} and is instead given by
$Q_\alpha^{\alpha\beta} = \frac{1}{2}\mb{R}_\alpha^{\alpha\beta}\cdot(\mb{v}_{\beta} - \mb{v}_{\alpha}) + 3\frac{m_{\alpha\beta}}{m_{\alpha}}n_{\alpha}\nu_{\alpha\beta}k_{B}(T_{\beta}-T_{\alpha})$.

\nind\textit{Ohm's Law:} \\
The generalized Ohm's law used in this paper which the viscous part of the electron pressure tensor but includes electron-neutral collisions as well as electron-ion collisions:
\be 
\mb{E} + (\mb{v}_i \times \mb{B}) =
\eta\mathbf{j} + \frac{\mb{j}\times\mathbf{B}}{en_{i}} - 
\frac{1}{en_{i}}\nabla P_{e} -
\frac{m_{e}\nu_{en}}{e}\mathbf{w},
\label{eqn:ohms}
\ee
where $\mb{w}=\mb{v}_{i}-\mb{v}_{n}$.
The resistivity $\eta$ is calculated using the electron-ion and electron-neutral collisional frequencies, and thus depends on the plasma conditions:
\be
\eta=\frac{m_{e}n_{e}(\nu_{ei}+\nu_{en})}{(e n_{e})^2}
\label{eqn:res}
\ee 
where the electron-ion ($\nu_{ei}$) and electron-neutral ($\nu_{en}$) collision frequencies are given by
\be
\nu_{e i} = \frac{4}{3}n_{i}\Sigma_{e i}\sqrt{\frac{2k_{B}T_{e}}{\pi m_{e}}}, ~ \textrm{and} ~
\nu_{e n} = n_{n}\Sigma_{e n}\sqrt{\frac{8k_{B}T_{en}}{\pi m_{e n}}}.
\ee
The electron-ion collisional cross-section $\Sigma_{e i} = \lambda \pi r_{d,e}^{2}$, where $r_{d,e}=e^{2}/(4\pi\epsilon_{0} k_{B}T_{e})$ is the distance of closest approach for electrons. The electron-neutral collisional cross-section is $\Sigma_{en} = \Sigma_{ne} = 1\times 10^{-19} ~ \textrm{m}^2$. 
The model used in Leake et al. 2012 \cite{Leake2012} neglected all but the first term on the right hand side of the generalized Ohm's law, Equation (\ref{eqn:ohms}), and used a parameter for the resistivity $\eta$, rather than calculating it using Equation (\ref{eqn:res}).

\nind\textit{Summary:} \\
The changes made to the model presented in Leake et al. 2012 \cite{Leake2012}, which have been implemented to make the physical model more realistic, have a number of consequences. In the model in this paper we include charge exchange interactions, which, for the chromospheric  plasma parameters, effectively doubles the collisional coupling between ions and neutrals. Also, the ion-neutral scattering cross-section is larger here than in Leake et al. 2012 \cite{Leake2012}, as we use a more realistic value, based on the theoretical work of Ref.  \onlinecite{1983ApJ...264..485D}. These two changes increase the neutral-ion collisional coupling, which reduces the neutral-ion mean free path by approximately an order of magnitude. As we will show in the following section, the Sweet-Parker width in this simulation is also smaller. As a result the ratio of Sweet-Parker width to neutral-ion collisional mean free path in the simulation in this paper is similar to that in Leake et al. 2012 \cite{Leake2012}.

In addition to the above changes, in the model used in this paper we evolve the system with a generalized Ohm's law where the resistivity is not a parameter of the simulations, but appropriately depends on the electron collision frequencies ($\nu_{ei},\nu_{en}$) and hence depends on the local plasma parameters. This is more realistic than performing a parameter study over resistivity as was done in Leake et al. 2012 \cite{Leake2012}, and helps to confirm that the results found in that paper are valid for realistic chromospheric parameters. In addition to these differences the simulations in this paper have a magnetic Prandtl number lower than the simulations of Leake et al. 2012 \cite{Leake2012}, as we are using a different resistivity and more realistic ion and neutral transport coefficients. This will be discussed in more detail  in the next section.

\subsection{Normalization}
The equations are non-dimensionalized 
by dividing each variable ($C$) by its normalizing value ($C_{0}$).
The set of equations requires a choice of three normalizing values. Normalizing values for the length ($L_{0}=1\times10^{4} ~ \textrm{m}$), number density ($n_{0}=3\times10^{16} ~ \textrm{m}^{-3}$), and magnetic field ($B_{0}=2\times10^{-3} ~ \textrm{T} $) are chosen. From these values the normalizing values for the velocity ($v_{0}=B_{0}/\sqrt{\mu_{0}m_{i}n_{0}}=2.52\times10^{5} ~ \textrm{m}.\textrm{s}^{-1}$), time ($t_{0}=L_{0}/v_{0} =0.04$ s), temperature ($T_{0}=B_{0}^{2}/k_{B}\mu_{0}n_{0} = 7.69\times10^{6}~\textrm{K}$), pressure ($P_{0}=B_{0}^2/\mu_{0}= 3.18 ~ \textrm{Pa}$), current density ($j_{0}=B_{0}/(\mu_{0}L_{0})=0.6 ~ \textrm{A}.\textrm{m}^{2}$), and resistivity ($\eta_{0} = \mu_{0}L_{0}v_{0} = 3.17\times10^{3} ~ \Omega$m) can be derived. 

\subsection{Initial conditions}

The simulation domain extends from -16$L_{0}$ to 16$L_{0}$ in the $x$ direction and -4$L_{0}$ to +4$L_{0}$ in the $y$ direction, with a periodic boundary condition in the $x$-direction and perfectly-conducting boundary conditions in the $y$-direction. The horizontal extent is chosen to be larger than the maximum horizontal extent of the current sheet length during the simulation (which is approximately $6L_{0}$).

The initial neutral fluid number density is 200$n_{0}$ ($6\times 10^{18} ~ \textrm{m}^{-3}$), and the ion fluid number density is $0.2 n_{0}$ ($0.6
\times 10^{16} ~ \textrm{m}^{-3}$). This gives a total (ion+neutral) mass
density of $1.0\times10^{-8} ~ \textrm{kg}.\textrm{m}^{-3}$, and an
initial ionization level ($\psi_{i}\equiv n_{i}/(n_{i}+n_{n}$)) of 0.1
\%. The initial electron, ion and neutral temperatures are set to $1.1\times 10^{-3}T_{0}$ ($8.46 \times 10^{3} ~ \textrm{K})$. These initial conditions are consistent with lower to middle chromospheric conditions, based on 1D semi-empirical models of the quiet Sun \citep{1981ApJS...45..635V}. For these plasma parameters, the isotropic neutral heat conduction is much faster than any of the anisotropic heat conduction tensor components for either ions or electrons.  This fact, coupled with the fact that the ion-neutral collision time is small compared with a typical dynamical time, means that thermal diffusion of the plasma is dominated by neutral heat conduction and is primarily isotropic.

As in Leake et al. 2012 \cite{Leake2012}, the initial magnetic configuration is a Harris current sheet, 
\be
A_{z} = - \frac{B_{0}}{2}\lambda_{\psi}\ln{\cosh{(y/\lambda_{\psi})}}, ~ \mb{B} = \nabla \wedge A_{z}\mb{\hat{e}}_{z}
\ee
with an initial width of $\lambda_{\psi}=0.5L_{0}$. Both the
ionized pressure $P_{p} =P_{i}+P_{e}$ and the neutral pressure
$P_{n}$ are increased in 
the current sheet in order to balance the Lorentz force of the magnetic field in the current sheet.
\bea
P_{p}(y) & = & P_{p} + \frac{1}{2}\frac{F}{\cosh^{2}{(y/\lambda_{\psi})}}, \\
P_{n}(y) & = & P_{n} + \frac{1}{2}\frac{P_{0}-F}{\cosh^2{(y/\lambda_{\psi})}},
\eea
where $F=2\times 10^{-3}P_{0}$ is chosen to maintain an approximate ionization
balance of 0.1\% inside the current sheet. 

A relative velocity between
ions and neutrals is therefore required so that the collisional terms (both scattering collisions and charge exchange) 
can couple the ionized and neutral fluids and keep the fluids individually in approximate force balance. 

For simplicity, using the fact that the ion-neutral drag due to charge exchange and scattering collisions are approximately equal, we assume the following form 
for the initial ion velocity
\be
{v_{i}}_{y}(y) = \frac{(F-1)}{n_i n_n\nu_{in}}\frac{\tanh{(y/\lambda_{\psi})}} {\lambda_{\psi}\cosh^{2}{(y/\lambda_{\psi})}}\frac{n_{0}^2v_{0}^2}{P_{0}}.
\ee
While this is not an exact steady state, the velocities that arise from the unbalanced forces in the initial condition
are very small compared with the flows created by the onset of the reconnection process.

To initiate the onset of the tearing mode instability, an additional, localized, small amplitude rotational flow perturbation is applied to the neutral and ion fluids, with the form
\bea
v_{\alpha,y} & = & -\frac{v_{\delta}}{5}^{0}\sin{\left(\frac{y}{\lambda_{\psi}}\right)}\left[ \cos{\left(\frac{x}{5\lambda_{\psi}}\right)}  - 2\frac{x}{5\lambda_{\psi}}\sin{\left(\frac{x}{5\lambda_{\psi}}\right)} 
\right]e^{-r^{2}} \\
v_{\alpha,x} & = & {v_{\delta}}^{0}\sin{\left(\frac{x}{5\lambda_{\psi}}\right)}\left[ \cos{\left(\frac{y}{\lambda_{\psi}}\right)}  - 2\frac{y}{\lambda_{\psi}}\sin{\left(\frac{y}{\lambda_{\psi}}\right)} 
\right]e^{-r^{2}}
\eea
where ${v_{\delta}}^{0} = 5\times 10^{-4}v_{0}$ and $r=\frac{ (\frac{x}{5})^{2}+ y^2}{\lambda_{\psi}^{2}}$. 

For these initial conditions, the background neutral-ion collisional mean free path, $\lambda_{ni}=v_{T,n}/\nu_{ni}$ is 90 m. The initial current sheet width is $0.5L_{0}$=
5 km, and so is much larger than $\lambda_{ni}$. Therefore we expect that the initial reconnection will behave like the classical single-fluid (i.e., when the neutrals and ions are coupled) Sweet-Parker model, and the initial current sheet will thin until it reaches the Sweet-Parker width, $\delta_{SP}$. To estimate this Sweet-Parker width, we use the equation $\delta_{SP} \approx \sqrt{\eta L/{\mu_{0}v_{A}}}$, use the initial resistivity $\eta$, background field strength and total number density (to get an estimate for $v_{A}$), and  estimate that the reconnection region length $L$ is equal to the spatial extent of the tearing perturbation ($L=5\lambda_{\psi}=2.5\times10^{4}$ m). This gives an estimate of $\delta_{SP}\approx 120$ m, 
which is comparable to the background neutral-ion collisional mean free path, $90$ m. If the current sheet thins until its width is comparable to the $\delta_{SP}$, its width will also be comparable to $\lambda_{ni}$, and so the neutrals will begin to decouple from the ions. The decoupling will result in the neutral pressure being unable to balance magnetic pressure and the sheet will thin further. Thus the current sheet width may decrease beyond the Sweet-Parker width, and eventually may become less then the mean free path.  This scenario occurred in the simulations of Leake et al. 2012 \cite{Leake2012}, although the physical model and initial parameters were different from the simulation in this paper. In Leake et al. 2012 \cite{Leake2012}, the background neutral-ion collisional mean free path was 1350 m, and the Sweet-Parker width had a range of 1330 m to 8410 m (as the resistivity $\eta$ was varied as a parameter of the simulations). The current sheets in those simulations thinned to less than 1000 m, i.e., narrower than the mean free path. 

\section{Results}

\begin{figure}
\begin{center}
\includegraphics[width=\textwidth]{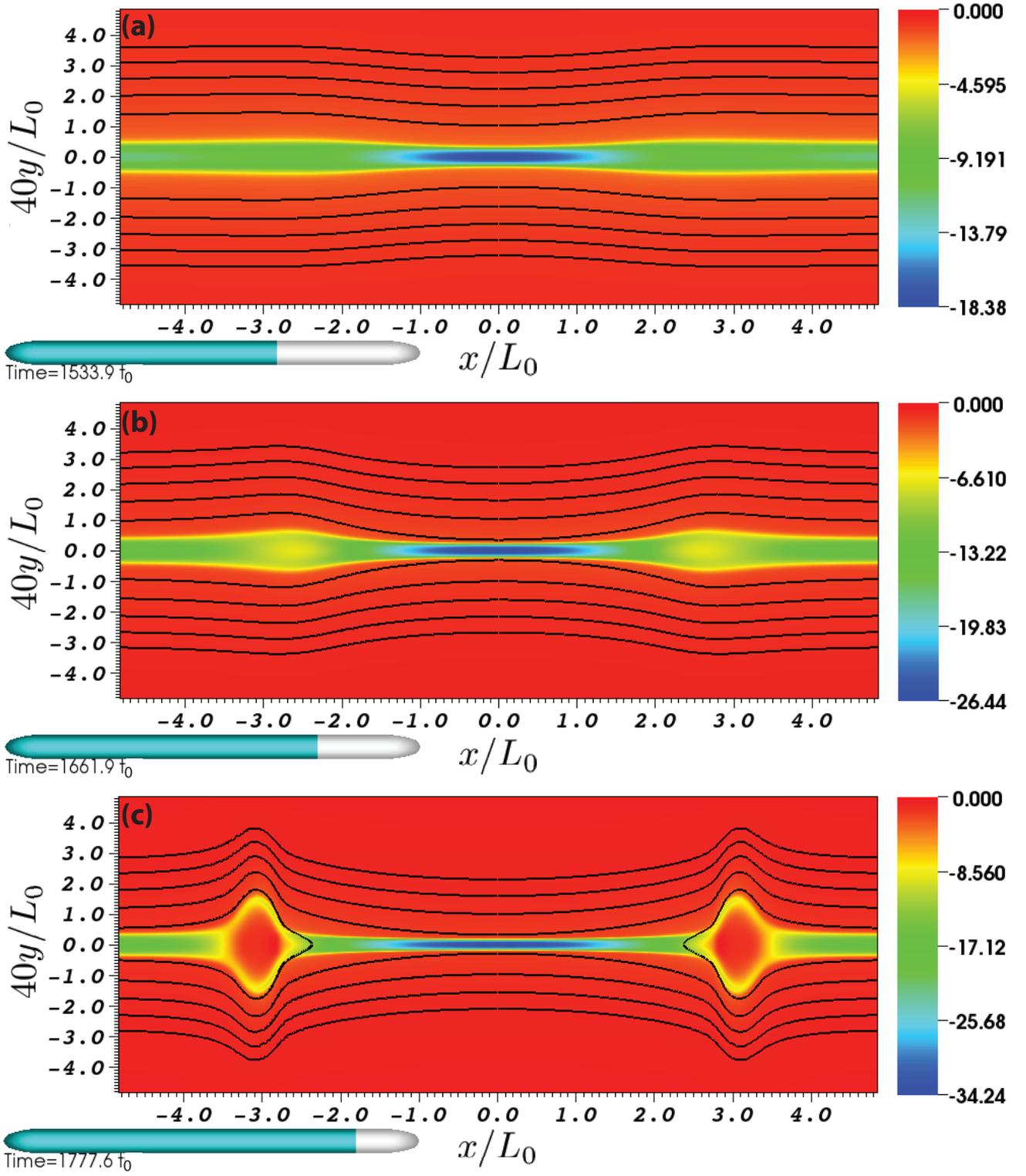}
\caption{Formation of reconnection region. Current density ($j_{z}/j_{0}$) is shown by color scale, and 5 contour values of $A_{z}$, $[-0.02,-0.0175, -0.015, -0.0125, -0.01]B_{0}L_{0}$, show representative fieldlines. Note that the $y$-coordinate is expanded  by a factor of 40.
\label{fig:reconnection}}
\end{center}
\end{figure}

Figure \ref{fig:reconnection} shows the early evolution of the reconnection region resulting from the initial perturbation, on a subset of the simulation domain. The color indicates current density, and the solid lines show isocontours of $A_{z}$, i.e., magnetic fieldlines.
The $y$ axis is stretched by a factor of 40 to show the thin structures formed. The tearing instability grows, magnetic field reconnects at the X-point, and flux is ejected. A Sweet-Parker-like reconnection region is formed, where  ions are pulled in by the magnetic field and drag the neutrals with them. 

\begin{figure}
\begin{center}
\includegraphics[width=\textwidth]{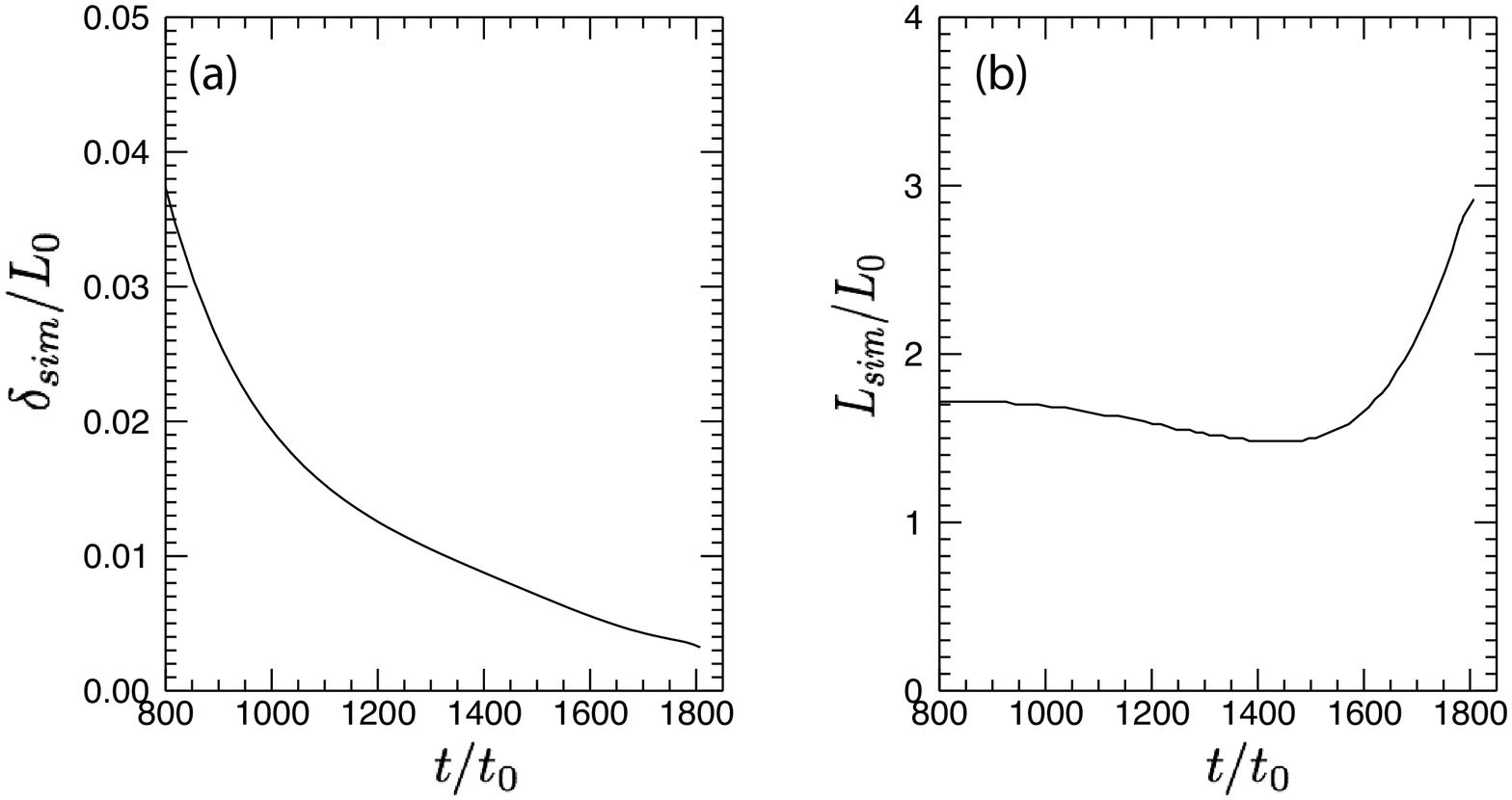}
\caption{Width and length of the current sheet as a function of time. The width is defined by the half-width half-maximum based on the current density $j_{z}$. The length is defined as the distance from $x=0$ to the $x$ location of maximum ion outflow $v_{i,y}$ on the $y=0$ line. The current sheet width monotonically decreases during the simulation after t=800 s, and eventually becomes less than the background neutral-ion collisional mean free path $\lambda_{ni}$, which is $0.0095L_{0}=95$ m at the end of the simulation. Before 800 s, the tearing instability has not developed enough to give a uniquely defined current sheet width and length.  As the non-linear reconnection continues, the current sheet is seen to elongate after t=1500 s. 
\label{fig:aspect_ratio}}
\end{center}
\end{figure}

The current sheet width and length as functions of time are shown in Figure \ref{fig:aspect_ratio}. The width, $\delta_{sim}$, is defined to be the half-width at half-max of the current sheet. The length, $L_{sim}$, is defined as the $x$ location along the line $y=0$ at which the horizontal velocity of the ions reaches a maximum (i.e., the location of maximum outflow). These two definitions are the same as was used in Leake et al. 2012 \cite{Leake2012}.  As can be seen in Figure \ref{fig:aspect_ratio}, the current sheet width falls from a value of $0.037 L_{0} = 370$ m at time $t=800t_{0}$ to a value of $3.3\times 10^{-3}L_{0}=33$ m by the end of the simulation. This final width is less than the neutral-ion collisional mean free path at this time ($95$ m). Thus during the simulation the current sheet has undergone a transition from being thicker than the mean free path to being thinner than it, as predicted. The current sheet also lengthens after $t=1500t_{0}$, increasing to  $2.9L_{0}$. The aspect ratio ($\delta_{sim}/L_{sim}$) at $t=1800t_{0}$ is $1.1\times 10^{-3}$.

\begin{figure}
\begin{center}
\includegraphics[width=\textwidth]{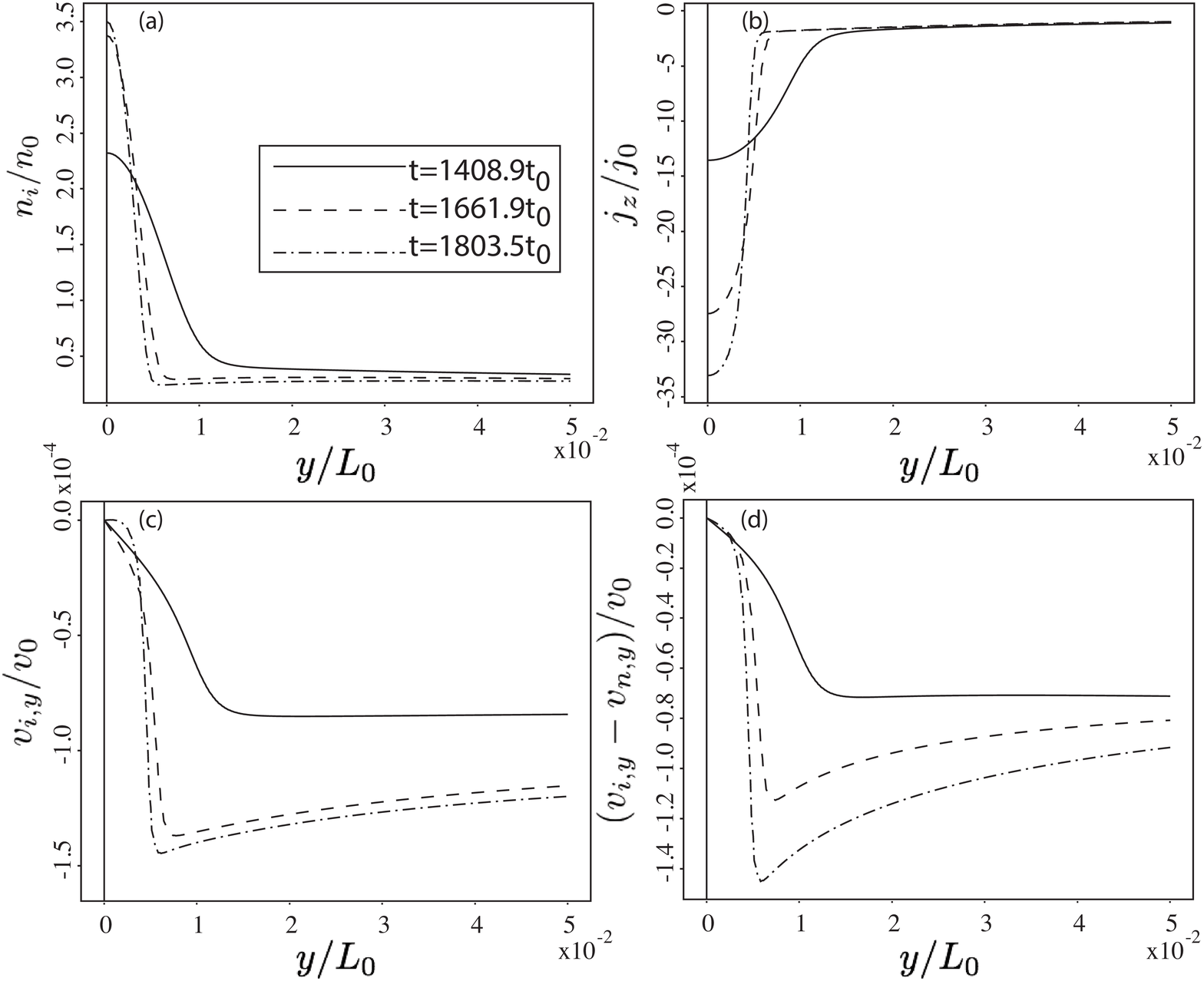}
\caption{
Profiles across the current sheet ($x=0$). Panel (a) shows the ion density ($n_{i}/n_{0}$), Panel (b) shows the current density ($j_{z}/j_{0}$), Panel (c) shows  the ion inflow ($v_{i,y}/v_{0}$), and Panel (d) shows the difference in neutral and ion inflow ($(v_{i,y}-v_{n,y})/v_{0}$). This figure demonstrates the sharpening in the current density and ion density as the current sheet collapses. It also indicates that the peak in $|v_{n,y}|$ is approximately 0.1 times the peak in $|v_{i,y}|$, i.e., the inflow is decoupled.
\label{fig:width}}
\end{center}
\end{figure}

Figure \ref{fig:width} shows profiles across the current sheet at $x=0$ of the ion density, current density, vertical ion flow, and the difference between the vertical neutral flow and the vertical ion flow, at three different times in the simulation (1408.9$t_{0}$, 1661.9$t_{0}$, and 1803.5$t_{0}$).
As the tearing instability develops, Panels (a) and (b) show the current density and ion density building into a sharp structure with the same gradient scale. Note that at $y=0$, the ionized fraction ($n_{i}/(n_{i}+n_{n})$) rises by an order of magnitude from 0.1\% at $t=0$ to 1.2\% at the end of the simulation ($t=1807t_{0}$), though the plasma remains weakly ionized throughout the domain. Panels (c) and (d) show an increase in the ion inflow and an increase in the difference between neutral and ion inflows as the current layer develops. The amplitude of the ion inflow reaches a maximum of $1.5\times10^{-4}v_{0}$, while the neutral inflow is much slower than the ion inflow. The peak difference between neutral and ion inflows at $t=1803.5$ s is $1.4\times10^{-4}v_{0}$, which means that the peak neutral inflow is  $1\times10^{-5}v_{0}$, less than a tenth of the peak ion inflow. Clearly the neutrals have become decoupled from the ions, and are being pulled into the reconnection region much slower than the ions.

\begin{figure}
\begin{center}
\includegraphics[width=\textwidth]{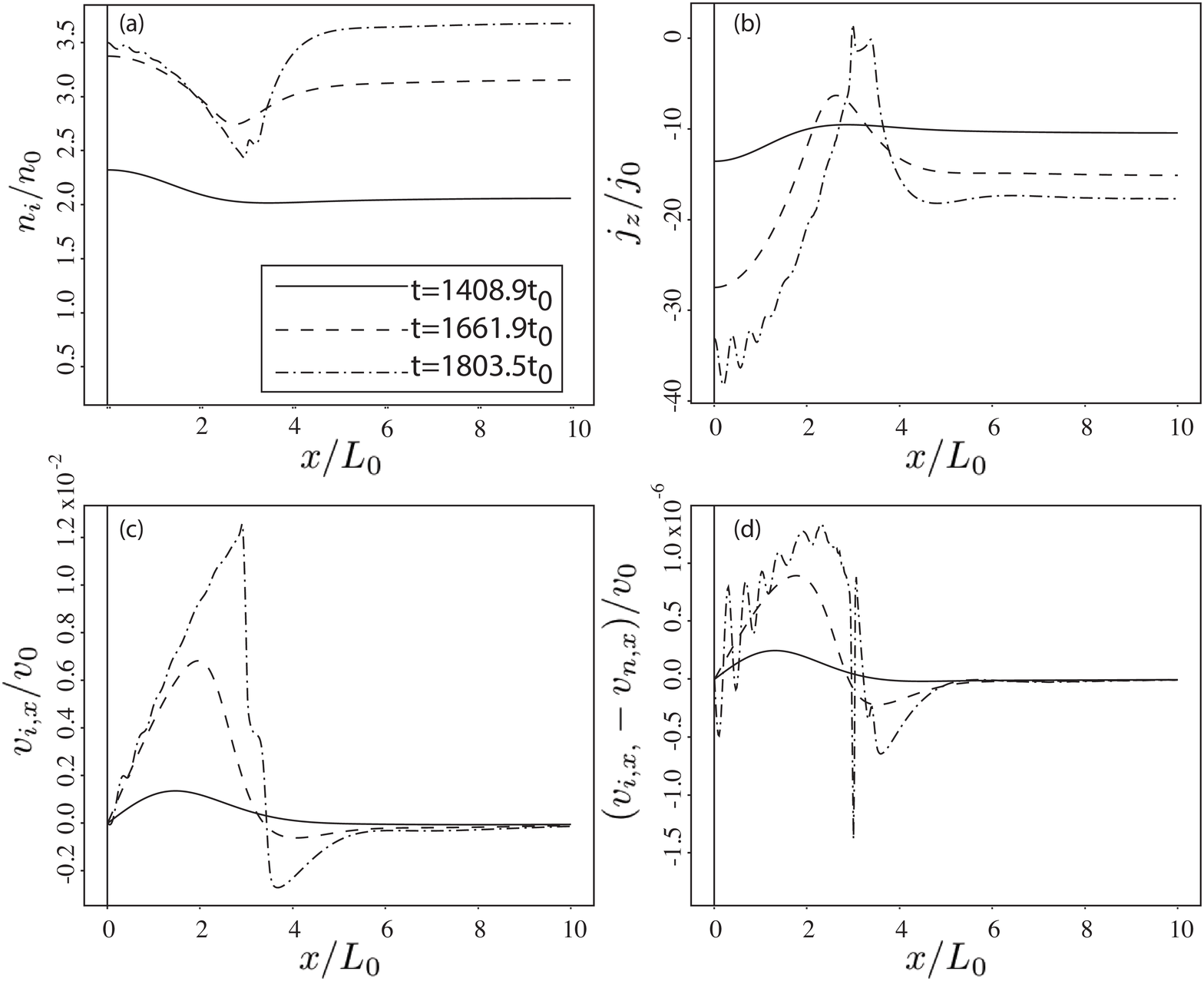}
\caption{
Profiles along the current sheet ($y=0$). Panel (a) shows the ion density ($n_{i}/n_{0}$). Panel (b) shows the current density ($j_{z}/j_{0}$). Panel (c) shows  the ion outflow ($v_{i,x}/v_{0}$). Panel (d) shows the difference in neutral and ion outflow ($(v_{i,x}-v_{n,x})/v_{0}$). The difference in ion and neutral outflow is small compared to the ion outflow, and so the neutrals and ions fluids are coupled. The $x$ location of the peak in the ion outflow can be seen to move outward in time (Panel (c)).
\label{fig:length}}
\end{center}
\end{figure}

Figure \ref{fig:length} shows profiles of the ion density, current density, ion outflow and the difference between the neutral and ion outflows, within the current sheet, along the line $y=0$ at three different times. The wave-like oscillations seen in the current density profile between $x=0$ and $x=L_{0}$ in Panel (b) are caused by the onset of the secondary tearing instability, which will be discussed later in this section. Also note the increase in ion density and moderate increase in current density magnitude outside the reconnection region, $x\gtrsim3.5L_{0}$, at later times. This behavior is characteristic of the 1D magnetic field reconnection in a weakly ionized plasma considered by Ref. \onlinecite{1999ApJ...511..193V} and Ref. \onlinecite{2003ApJ...583..229H}. As discussed in the above referenced studies, such a recombination facilitated process can lead to resistivity independent reconnection for a limited range of plasma parameters.  However, as will be shown here and as was shown by Leake et al. 2012 \cite{Leake2012}, by allowing for plasma outflow in a 2D configuration and with given plasma conditions, the reconnection rate can be accelerated beyond the 1D result.

\begin{figure}
\begin{center}
\vspace{-5mm}
\includegraphics[width=0.95\textwidth]{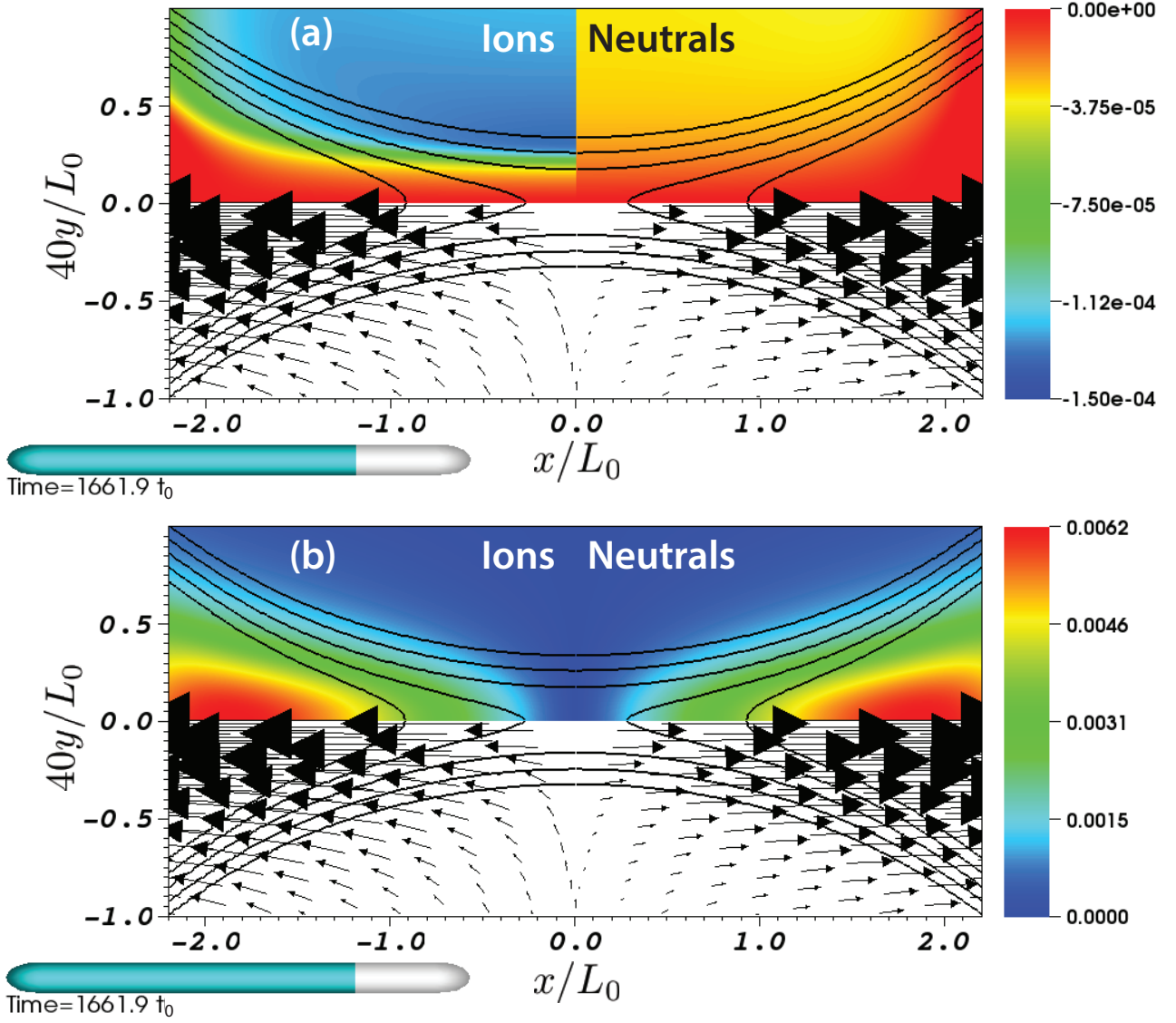}
\vspace{-0mm}
\caption{Illustration of the decoupling of ion and neutral inflow during reconnection. At $t=1661.9t_{0}$, the current sheet width $\delta_{sim}$ is less than the background neutral-ion collisional mean free path $\lambda_{ni}$. Panel (a): The ion (top left quadrant) and neutral (top right quadrant) inflows ($v_{i,y}/v_{0}$ and $v_{n,y}/v_{0}$). The solid lines show 5 contour values of $A_{z}$, $[-0.01, -0.00975, -0.0095, -0.00925, -0.009]B_{0}L_{0}$. The vectors show ion flow on the bottom left quadrant and neutral flow on the bottom right quadrant. 
Panel (b): The ion and neutral outflows ($v_{i,x}/v_{0}$ and $v_{n,x}/v_{0}$):  The contour lines and vectors are the same as in the top panel.
\label{fig:velocities}}
\end{center}
\end{figure}

Figure \ref{fig:length}, Panels, (c) and (d), show that the neutral and ion outflow are very well coupled with the difference between neutral and ion outflow being approximately 4 orders of magnitude smaller than the ion outflow.  To highlight the fact that the neutral inflow is decoupled from the ion inflow, but that the outflows are coupled,
Figure \ref{fig:velocities}(a) shows a subset of the simulation domain at $t=1661.9t_{0}$. The solid lines show isocontours of $A_{z}$. The top left quadrant shows 
the ion inflow velocity, and the top right quadrant shows the neutral inflow velocity. At this time, the current sheet width $\delta_{sim}=0.0045L_{0}=45$ m, which is smaller than the neutral-ion collisional mean free path $\lambda_{ni}=0.0095L_{0}=95$ m. The ions have a larger maximum amplitude of inflow than the neutrals, which is a direct consequence of the fact that the current sheet is now narrower than the neutral-ion collisional mean free path and the neutrals are not dragged into the current sheet as quickly as the ions.
Figure \ref{fig:velocities}(b) shows the outflow velocities. The outflows are well coupled in the sense that the difference between ion and neutral outflow is negligible compared to the magnitude of the ion outflow. This situation, where the inflow is decoupled  but the outflow is coupled, was also observed in the earlier simulations by Leake et al. 2012 \cite{Leake2012}, and is a consequence of the neutral-ion mean free path lying between the inflow and outflow scales of the current sheet, $\delta_{sim}< \lambda_{ni} < L_{sim}$.

The maximum outflow speed during the simulation is $0.0126v_{0}=3 ~ \textrm{km}/\textrm{s}$ which, although small compared to typically measured spicule flows (up to 150 km/s), is close to the flows observed in chromospheric jets ($\sim 10$ km/s). The Alfv\'{e}n speed just upstream of the reconnection current sheet is $0.006v_{0}$, and so the outflow can be up to twice the local upstream Alfv\'{e}n speed. As the Alfv\'{e}n speed scales with magnetic field strength and density, it is possible that larger outflows can be obtained by using either larger fields ($B_{0}=20$ G here), corresponding to more active regions of the Sun, or by using lower number densities, corresponding to conditions in the upper chromosphere.

\begin{figure}
\begin{center}
\includegraphics[width=\textwidth]{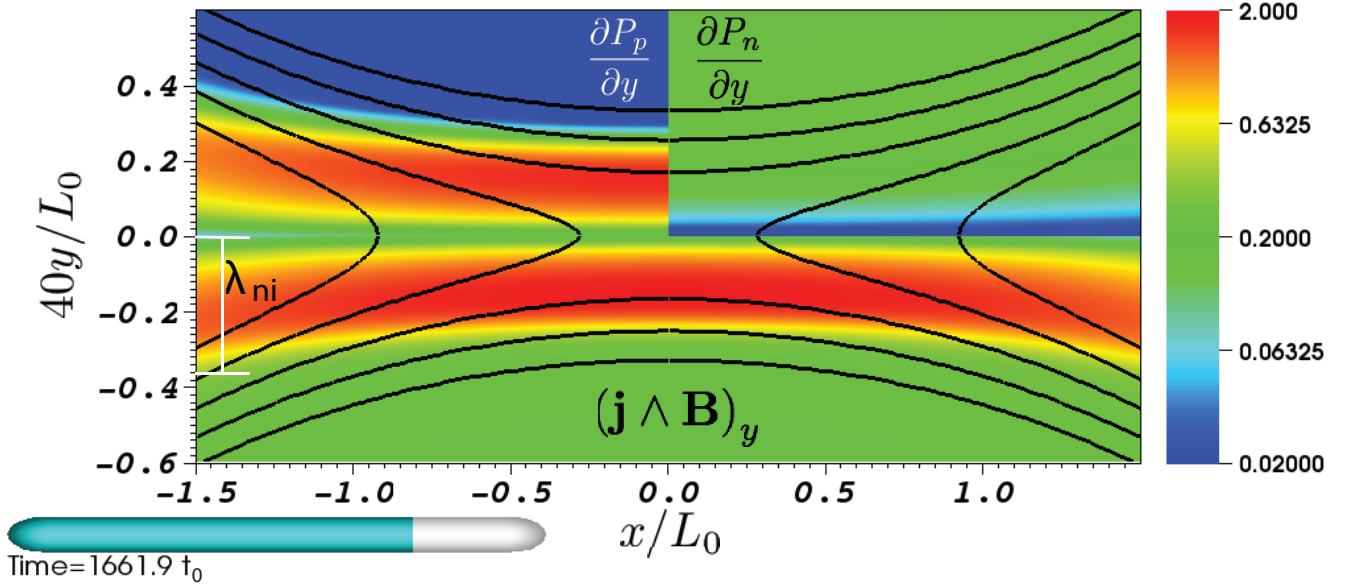}
\caption{
Contributions to the plasma and neutral momentum equations. The top left quadrant shows $\frac{\partial P_{p}}{\partial y}$ where $P_{p}=(P_{i}+P_{e})$, and the top right quadrant shows $\frac{\partial P_{n}}{\partial y}$. The bottom half shows ${(j\wedge B)}_{y}$. The color is on a log-scale. This shows that  $\frac{\partial P_{p}}{\partial y}$ balances ${(j\wedge B)}_{y}$ on the collisional scale $\lambda_{ni}$, while  $\frac{\partial P_{n}}{\partial y}$ balances ${(j\wedge B)}_{y}$ on scales greater than $\lambda_{ni}$. The solid lines are 5 contour values of $A_{z}$, $[-0.01, -0.00975, -0.0095, -0.00925, -0.009]B_{0}L_{0}$.
 \label{fig:R_in}}
\end{center}
\end{figure}

Figure \ref{fig:R_in} displays contributing terms to the $y$ component of the total momentum equation at time $t=1661.9t_{0}$.   The top left quadrant shows the vertical gradient of the ionized (ion+electron) pressure,  and the top right quadrant shows the vertical gradient in the neutral pressure. The bottom half shows the $y$ component of the Lorentz force. The color scale highlights that on the scale of $\lambda_{ni}$, it is 
the ionized pressure that balances the Lorentz force (in the $y$ direction), and on scales larger than $\lambda_{ni}$ it is the neutral pressure that balance the Lorentz force.

\begin{figure}
\begin{center}
\includegraphics[width=\textwidth]{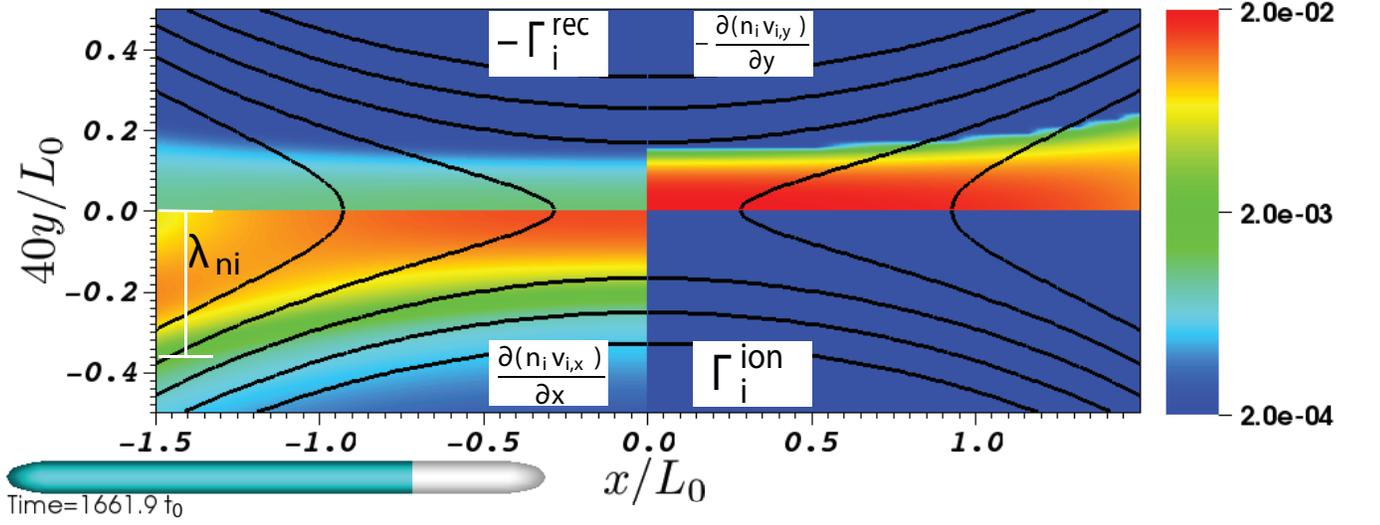}
\caption{The steady state reconnection region showing contributing sources and sinks of ions in the current sheet (in units of $L_{0}^{-3}t_{0}^{-1}$): Top left quadrant shows rate of loss of ions due to recombination. Bottom left quadrant shows rate of loss of ions due to outflow $\frac{\partial n_{i}v_{i,x}}{\partial x}$. Top right quadrant shows rate of gain of ions due to inflow  $-\frac{\partial n_{i}v_{i,y}}{\partial y}$. Bottom right quadrant shows rate of gain of ions due to ionization. The solid lines are 5 contour values of $A_{z}$, $[-0.01, -0.00975, -0.0095, -0.00925, -0.009]B_{0}L_{0}$. This shows that ionization is negligible, and that outflow is larger than recombination. \label{fig:steady_state}}
\end{center}
\end{figure}

\begin{figure}
\begin{center}
\includegraphics[width=\textwidth]{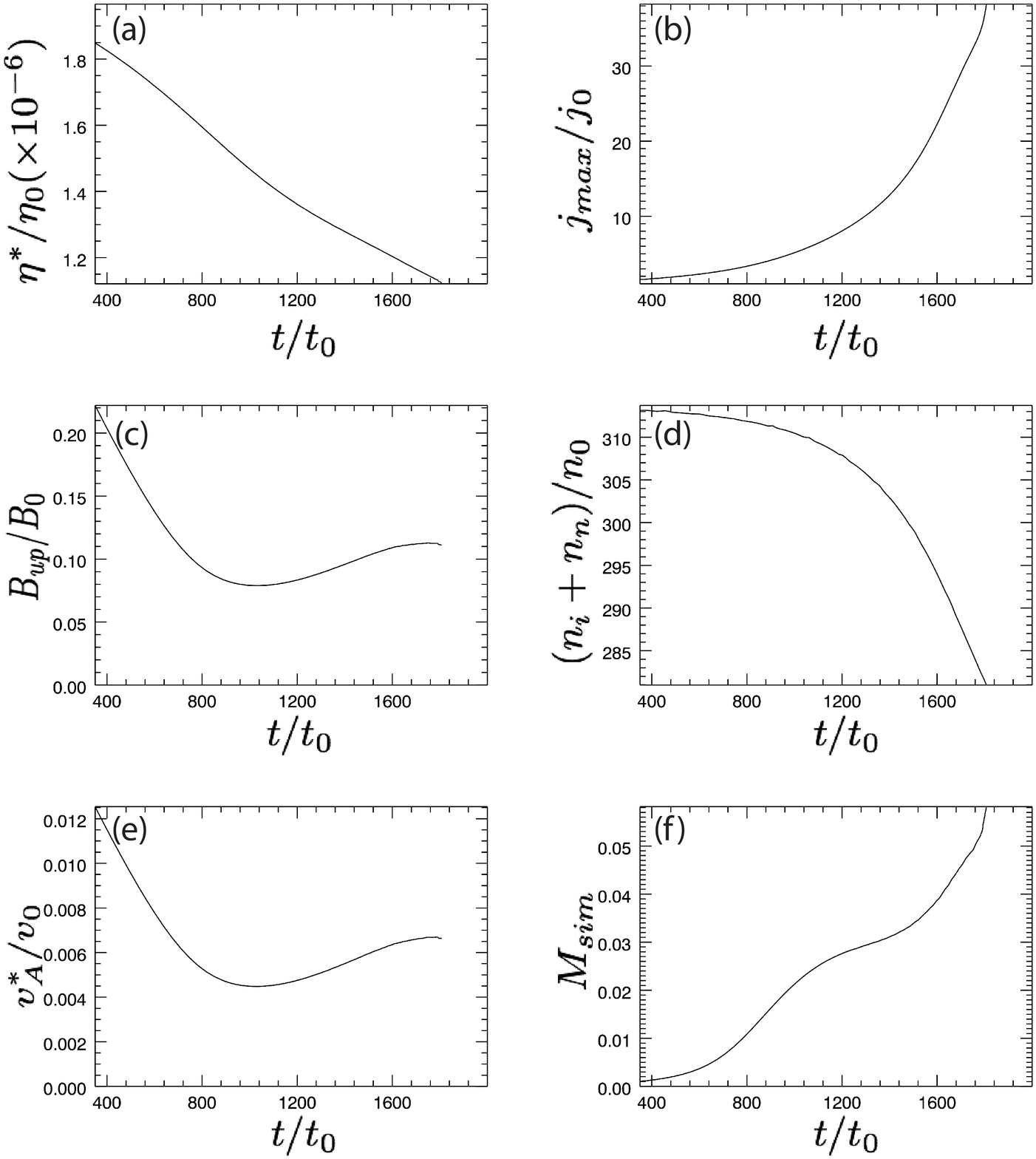}
\caption{Temporal evolution of the reconnection rate. Panel (a): Effective resistivity $\eta^{*}/\eta_{0}$. Panel (b): $j_{max}/j_{0}$, the maximum value of the current density, located at 
 $(x,y)=(x_{j},0)$, within the
reconnection region. Panel (c): $B_{up}/B_{0}$ where $B_{up}$ is 
evaluated at $(x_{j},\delta_{sim})$ and $\delta_{sim}$ is the half-width at half-max of the current sheet. Panel (d): $n^{*}/n_{0}$, the total density inside the current sheet at $(x_{j},0)$. Panel (e): $v_{A}^{*}/v_{0}$, the current sheet Alfv\'{e}n speed. Panel (f): $M_{sim} = \eta^{*} j_{max}/v_{A}^{*}B_{up}$. 
\label{fig:rec_rate}}
\end{center}
\end{figure}

The separation of inflow velocities has consequences for the reconnection occurring in the current sheet. If ions are pulled in faster than neutrals, with equal outflows, then there will be an excess of ions above the ionization equilibrium, and recombination will occur. 
In Leake et al. 2012 \cite{Leake2012} we found that recombination was as efficient at removing ions from the reconnection region as the outflow caused by reconnecting field. Figure \ref{fig:steady_state} shows the four components contributing
 to $\frac{\partial n_i}{\partial t}$ in the ion continuity
 equation for the simulation described here. The top left quadrant shows the loss due to recombination,
 the bottom left quadrant shows the  loss due to outflow (horizontal
 gradient in horizontal momentum of ions), the  top right quadrant
 shows the gain due to inflow (vertical gradient in vertical
 momentum), and the bottom right quadrant shows the gain due to ionization. Unlike in Leake et al. 2012 \cite{Leake2012}, here we observe that, within the reconnection
 region, outflow is larger than recombination, with a negligible contribution from ionization. An important distinction between the simulation in this paper and those in Leake et al. 2012 \cite{Leake2012}, is that in the present simulation the current sheet is never in steady state.  This is indicated by the monotonic decrease in current sheet width throughout the simulation, as shown in Figure \ref{fig:aspect_ratio}, Panel (a), as well as the temporal evolution of 
the current sheet, shown in Figure \ref{fig:rec_rate}. Hence the steady state analysis, which leads to the reconnection rate being determined by a balance between outflow, ionization and recombination, is not strictly valid. 

We now look at the temporal evolution of the reconnection rate in the current sheet. The normalized reconnection rate is defined by
\be
M_{sim} \equiv \frac{\eta^{*} j_{max}}{v_{A}^{*}B_{up}}.
\ee
Here, $j_{max}$ is the maximum value of the out of plane current density, $j_{z}$, located at $(x,y)=(x_{j},0)$. $B_{up}$ is $B_{x}$
evaluated at $(x_{j},\delta_{sim})$, where $\delta_{sim}$ has already been defined as the half-width half-max in out of plane current density.  $v_{A}^{*}$ is the relevant Alfven velocity defined using $B_{up}$ and the total number density ($n^{*}$) at the location of $j_{max}$. To obtain a local estimate of resistivity we use 
\be
\eta^{*} = \frac{(\mb{R}_{e}^{ei}+\mb{R}_{e}^{en})\cdot\mb{\hat{z}}}{en^{*}j_{max}}
\ee
with $\mb{R}_{e}^{ei}+\mb{R}_{e}^{en}$ evaluated at the location of $j_{max}$. The six quantities $\eta^{*}$, $j_{max}$, $n^{*}$, $B_{up}$, $v_{A}^{*}$, and $M_{sim}$ are shown as functions of time in Figure \ref{fig:rec_rate}.
The resistivity $\eta^{*}$ has two contributions, one from electron-ion collisions and the other from electron-neutral collisions. Taking the ratio of these two terms gives
\be
\frac{\nu_{en}}{\nu_{ei}} \approx \frac{3n_{n}}{2n_{i}}\frac{\Sigma_{en}}{\Sigma_{ei}}
\ee
where $\Sigma_{en}=1.16\times10^{-19} ~ \textrm{m}^{2}$, and  $\Sigma_{ei}$ depends on electron temperature as was described in Section 2.1. The ion, electron, and neutral temperatures have been assumed equal to each other in this estimate.
For the initial temperature of 8460 K, $\Sigma_{ei}=1.2\times 10^{-16} ~ \textrm{m}^{2}$, and at $t=0$, $n_{n}/n_{i}=10^3$. This gives a ratio of  $\frac{\nu_{en}}{\nu_{ei}}=1.5$. At  $t=1780t_{0}$,  the electron temperature is 9230 K,  $\Sigma_{ei}=1\times10^{-16} ~ \textrm{m}^{2}$ and $n_{n}/n_{i}=80$, which gives a ratio of  $\frac{\nu_{en}}{\nu_{ei}}=0.13$. Thus initially the electron-neutral contribution is of the same order as the electron-ion contribution, but toward the end of the simulation, the electron-ion contribution dominates due to the increase in ion density.

 The increase of the current density with time is primarily responsible for the monotonic increase of $M_{sim}$, from 0.001 at $t=400t_{0}$ to 0.056 at the end of the simulation. The Sweet-Parker model for magnetic reconnection predicts the normalized reconnection rate $M = 1/\sqrt{S}$ where $S$ is the Lundquist number of the reconnecting current sheet. The Lundquist number in this simulation, $S_{sim}$, is defined by
\be
S_{sim} \equiv \frac{v_{A}^{*}L_{sim}^{*}}{\mu_{0}\eta^{*}}.
\ee
At $t=1750$ s, before the plasmoid instability has developed, $\eta^{*}=1.205\times 10^{-6}\eta_{0}$, $L_{sim}^{*}=2.5L_{0}$, and $v_{A}^{*}=0.0066v_{0}$, which gives a 
value of $S_{sim}=1.37\times10^{4}$. Hence the Sweet-Parker model would predict a reconnection rate of $1/\sqrt{1.7\times10^{4}}=0.0085$. In this simulation we see the reconnection rate rise from below 0.001 to above 0.048 at $t=1750$ s. As the current sheet  width becomes less than $\lambda_{ni}$ and the neutrals decouple from the ions, it is clear that a single-fluid Sweet-Parker theory is no longer applicable. The reconnection rates observed in this simulation are sufficient to explain the observed lifetimes and reconnection rates of spicules, but not chromospheric jets.

\begin{figure}
\begin{center}
\includegraphics[width=\textwidth]{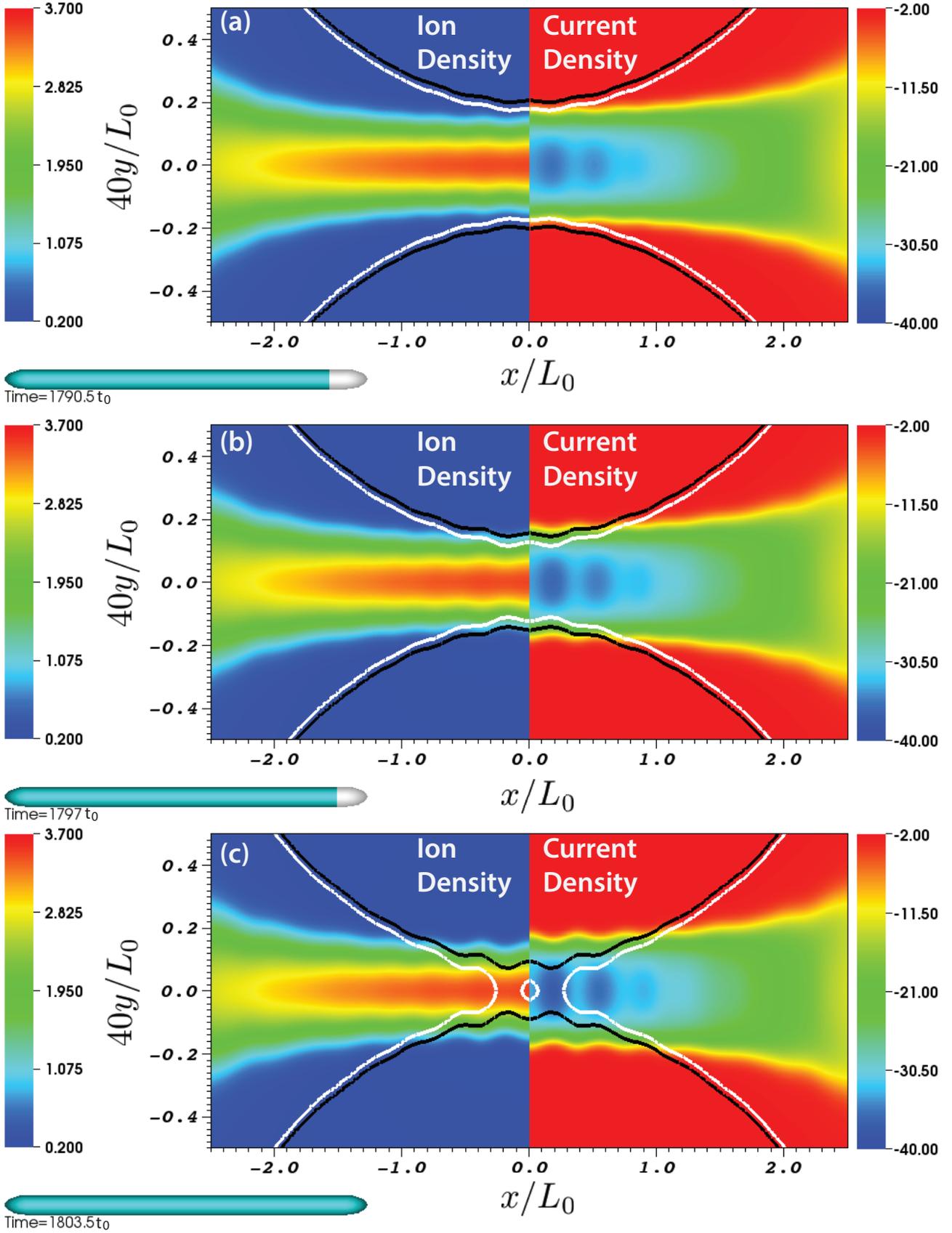}
\caption{Formation of plasmoids. Current density ($j_{z}/j_{0}$) on the right, and ion density $n_{i}/n_{0}$ on the left. The black line is the $A_{z}=-0.01245B_{0}L_{0}$ contour and the white line is the $A_{z}=-0.01237B_{0}L_{0}$ contour. Note that the $y$-coordinate is expanded  by a factor of 40.
\label{fig:plasmoids}}
\end{center}
\end{figure}

At $t=1790t_{0}$, the current sheet becomes unstable to the secondary tearing mode, known as the plasmoid instability (e.g., Ref. \onlinecite{Loureiro07}). Figure \ref{fig:plasmoids} shows the ion and current density, and selected isocontours of $A_{z}$. A plasmoid can be seen at the origin at $t=1803.5t_{0}$. The laminar current sheet can also be seen to break up into thinner current sheets. In the context of a single-fluid theory, Ref. \onlinecite{Loureiro07} predicted that the onset of the plasmoid instability occurs when the aspect ratio decreases to $1/200$ and this was the case in the multi-fluid simulations of Leake et al. 2012 \cite{Leake2012}. At $t=1700t_{0}$ the aspect ratio $\sigma_{sim}\equiv \delta_{sim}/L_{sim}$ already is 1/500, yet the plasmoids are not observed until $t=1790t_{0}$ when the aspect ratio has fallen further still. At this time we do not have a thorough understanding of why the onset of the plasmoid instability is different between the present simulation and our previous work. However, we note  that the magnetic Prandtl number, defined by the ratio of viscous to resistive diffusivity,
\be
Pr = \frac{\xi_{i}/(m_{i}n_{0})}{\eta/\mu_{0}},
\ee
 in this simulation is approximately 0.4. This is at least an order of magnitude lower than that used in Leake et al. 2012 \cite{Leake2012}, where $Pr$ ranged from 5 for high resistivity to 200 for low resistivity. The lower $Pr$ in the simulation in this paper means that viscous diffusivity is less important and this allows for stronger shearing of the reconnection outflows that may lead to stabilization of the secondary tearing instability \citep{1978JETPL..28..177B}. A detailed investigation of the stability of  weakly ionized reconnection current sheets to secondary tearing instability is left to future work.

\begin{figure}
\begin{center}
\includegraphics[width=\textwidth]{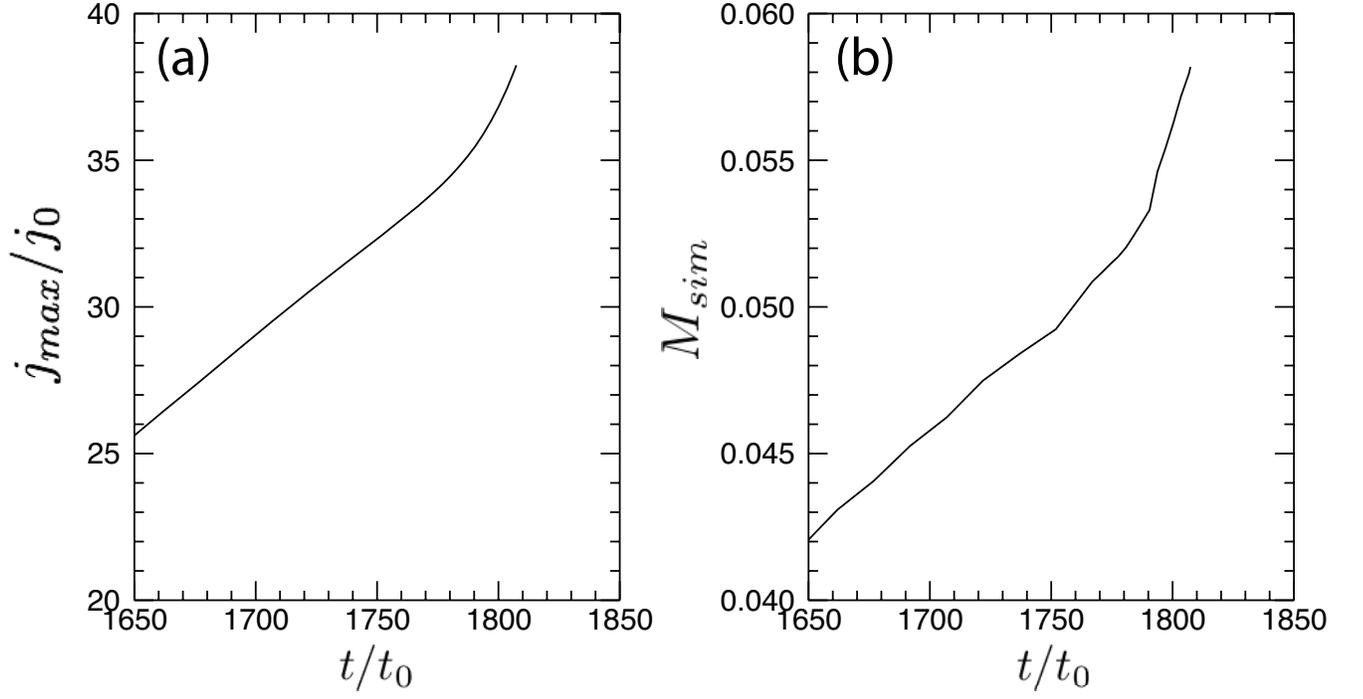}
\caption{Temporal evolution of the reconnection rate during the latter stages when the plasmoid instability sets in. Left panel: $j_{max}/j_{0}$, the maximum value of the out of plane current density. Right panel: $M_{sim} = \eta j_{max}/v_{A}^{*}B_{up}$, the normalized reconnection rate. 
\label{fig:plasmoids_rec_rate}}
\end{center}
\end{figure}

Figure \ref{fig:plasmoids_rec_rate} shows the reconnection rate $M_{sim}$ and $j_{max}$ in the interval \\
$t=[1650,1803.5]t_{0}$, i.e. a zoom in of Figure \ref{fig:rec_rate}, Panels (b) and (f). At $t=1790t_{0}$, there is a change in the time derivative of $j_{max}$ as the plasmoids create thinner current sheets and larger currents. This creates an increase in the reconnection rate. The further study of the non-linear evolution of the plasmoids is limited in this investigation, as the current sheet width approaches the limit of the resolution of the numerical simulation. 

In this paper magnetic reconnection in the solar chromosphere is simulated using a partially ionized reacting multi-fluid plasma model, which is able to capture the physical effects of decoupling between the ion and neutral fluids, occurring when scales associated with the reconnection region become comparable to the neutral-ion collisional mean free path. The number densities, magnetic field strengths, ionization levels and temperature are consistent with lower to middle chromospheric conditions in a quiet Sun region.

 These simulations improve upon the earlier work of Leake et al. 2012 \cite{Leake2012} by including the effects of charge-exchange interactions, and using more realistic values for the cross-sections of scattering collisions and transport coefficients.  In addition, the resistivity $\eta$ depends on the local plasma parameters, rather than being a fixed value. Even so, our simulation lies in the same regime as those of Leake et al. 2012 \cite{Leake2012}. This regime consists of an initial current sheet width $\delta_{sim}$ which is much larger than the Sweet-Parker width $\delta_{SP}$, which itself is comparable to the neutral-ion collisional mean free path:
 \be
  {\delta_{sim}}_{(t=0)} \gg \delta_{SP} \sim \lambda_{ni}.
  \ee
  In this regime, the current sheet width tends towards the Sweet-Parker width $\delta_{SP}$ (and hence towards $\lambda_{ni}$). When the current sheet width becomes comparable to the neutral-ion mean free path, decoupling of neutrals from ions allows the sheet width to fall below both the Sweet-Parker width and the mean free path:
  \be
 \delta_{SP},\lambda_{ni} > \delta_{sim}.
  \ee
  Once the current sheet width is smaller than the neutral-ion collisional mean free path, the steady state reconnection rate is determined by a combination of ion outflow and recombination. In the simulations of Leake et al. 2012 \cite{Leake2012}, recombination was found to be as important as outflow at removing ions from the reconnection region, and steady state reconnection rates faster than the Sweet-Parker prediction were seen. On the other hand, in the simulation presented in this paper no steady state is achieved. Instead we find an instantaneous reconnection rate which is faster than the Sweet-Parker prediction due primarily to the   separation of the current sheet width and the neutral-ion collisional scale and the resulting order of magnitude enhancement of the ion density in the current sheet with respect to the upstream value.
  
\begin{figure}
\begin{center}
\includegraphics[width=\textwidth]{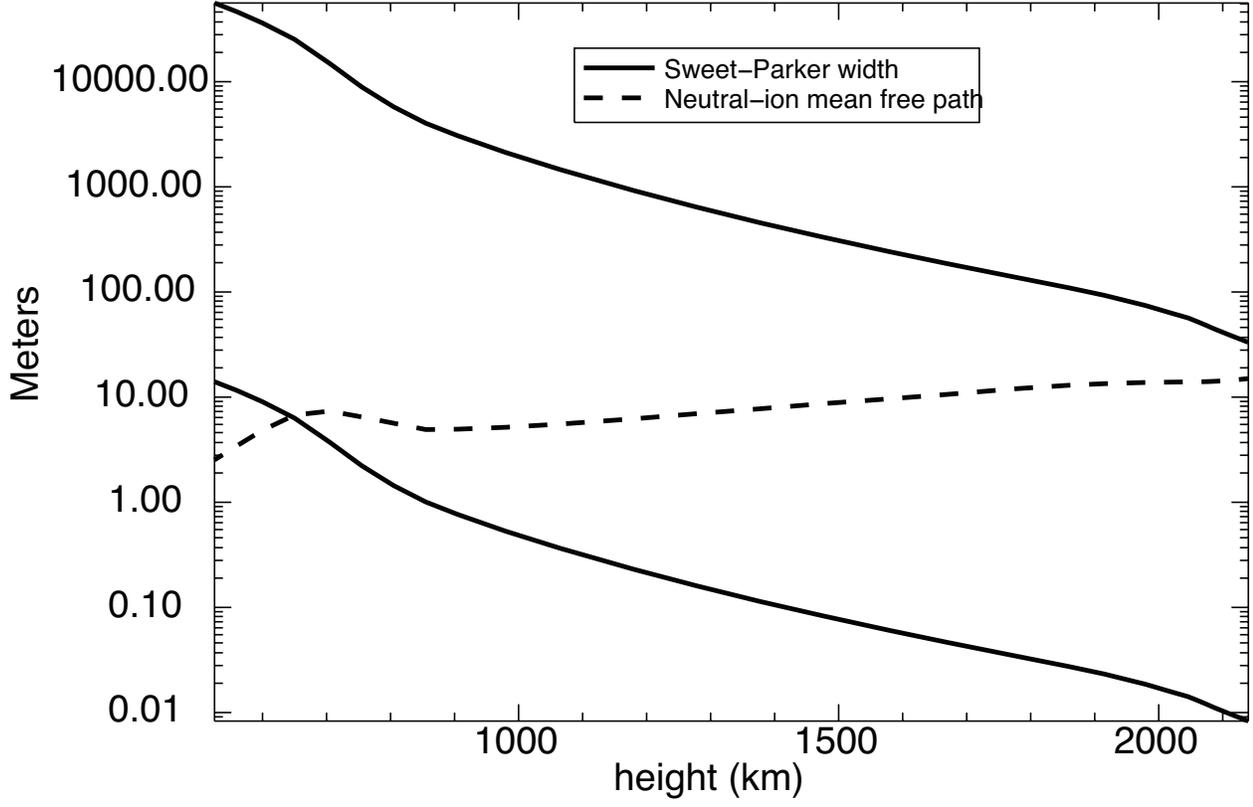}
\caption{Estimates of the Sweet-Parker width ($\delta_{SP}$) and neutral ion collisional mean free path ($\lambda_{ni}$) using the FALC model of the solar chromosphere \citep{2006ApJ...639..441F}. The solid lines show two extremes of the calculations of the resistive length. The higher value uses an aspect ratio of $\sigma=1/1000$ and a field strength of 5 G, and the smaller uses a current sheet aspect ratio of $\sigma=1/50$ and a field strength of 1000 G. The dot-dashed line is the neutral-ion collisional mean free path, $\lambda_{ni}$. This figure shows that for a range of heights in the chromosphere, the Sweet-Parker width can be larger than, but comparable to, the neutral-ion collisional mean free path, $\lambda_{ni}$.
\label{fig:lengths}}
\end{center}
\end{figure}

An important characteristic of the solar chromosphere is that it has a high degree of spatial variation of the plasma temperature and density due to stratification and horizontal structure, and also a high degree of spatial variation of the small scale magnetic field. This situation does not easily lend itself to investigating phenomena in a ``typical'' chromosphere. In particular, it is not clear that the regime we find in this simulation, where $\delta_{sim} < \lambda_{ni}$, is a ubiquitous occurrence in the chromosphere. To address this issue, we can estimate the altitude profiles of both the Sweet-Parker width and neutral-ion collisional mean free path using a semi-empirical 1D model of the quiet Sun \citep{2006ApJ...639..441F}, and a few assumptions.
The neutral-ion collisional mean free path depends on temperature and ion density via the thermal speed and neutral-ion collisional frequency. The temperature and density are provided by the 1D atmospheric model of Ref. \onlinecite{2006ApJ...639..441F}.
To obtain an altitude profile for the Sweet-Parker width we can replace the characteristic system size L with $\delta_{SP}/\sigma_{SP}$ in the equation $\delta_{SP}\approx\sqrt{\eta L /v_{A}\mu_{0}}$ to obtain
\be
\delta_{SP}\approx\frac{\eta}{\sigma_{SP}v_{A}\mu_{0}}.
\ee
Taking an aspect ratio at which the plasmoid instability should theoretically set in gives a lower bound on the Sweet-Parker aspect ratio $\sigma_{SP}$ and hence an upper bound on the Sweet-Parker width $\delta_{SP}$. This critical value varies in the literature, and so we take two extremes, 1/1000 (of the order of the smallest value seen in the simulation in this paper), and 1/50 (values observed in high plasma-beta simulations by Ref. \onlinecite{2012PhPl...19g2902N}).
An estimate of the magnetic field strength in the chromosphere is also required to estimate $\delta_{SP}$ in the chromosphere. We use the range [5,1000] G.
Using these two sets of extremes (one for $\sigma_{SP}$ and one for magnetic field strength) gives a range of altitude profiles for $\delta_{SP}$. Figure \ref{fig:lengths} depicts the two extreme profiles of $\delta_{SP}$, as well as the altitude profile of the neutral-ion collisional mean free path. This plot shows that for the ranges of aspect ratios and magnetic field strengths considered here, the Sweet-Parker width can be greater than, but comparable to, the neutral-ion collisional mean free path in a height range of 650 km to 2000 km above the solar surface. This essentially encompasses the entire chromosphere. Hence, the regime that our simulation self-consistently produces, where the reconnection transitions from being fully coupled to decoupled, is possible for a wide range of chromospheric conditions. This transition from coupled to decoupled reconnection is accompanied by  an increase in the reconnection rate.   

Recombination-aided fast reconnection in a weakly ionized plasma that is out of ionization balance was considered by Ref. \onlinecite{1999ApJ...511..193V} and Ref. \onlinecite{2003ApJ...583..229H}. Those theoretical investigations modeled the reconnection region as 1D and time independent with a width which was already less than the neutral-ion collisional mean free path (denoted $L_{AD}$ in Ref. \onlinecite{2003ApJ...583..229H}, and denoted $\lambda_{ni}$ in this paper). Hence the reconnection region was assumed to already have an excess of ions due to the decoupling of neutrals and ions. The simulation performed in this paper shows that for initial conditions relevant to the solar chromosphere, and a current sheet initially thicker than $\lambda_{ni}$, the system can self-consistently evolve to a current sheet width thinner than $\lambda_{ni}$. However, in this simulation, the recombination is less important than the outflow of ions, whereas in the 1D models of Ref. \onlinecite{1999ApJ...511..193V} and Ref. \onlinecite{2003ApJ...583..229H}, the recombination was assumed to be the only sink for the ions in the reconnection region.

As in Leake et al. 2012 \cite{Leake2012}, we measure ``fast'' reconnection rates close to $M=0.1$. Generally, fast reconnection requires either Hall, kinetic, or localized resistivity effects \citep{Birn01}, or secondary tearing \citep{Loureiro07,Huang10}. Hall effects are negligible in this simulation as the ion inertial scale is
 less than the smallest current sheet width in this simulation. In addition, the onset of the secondary tearing mode only increases the reconnection rate after $t=1750$ by approximately 15\%. In this simulation we obtain normalized reconnection rates above 0.05 without either Hall effects or secondary tearing. Given that we also have neither a localized increase in resistivity in the reconnection site, nor kinetic effects, we can conclude that fast reconnection is obtained solely due to the decoupling of neutrals from ions when the current sheet width is less than the neutral-ion collisional mean free path. Note, however, that Figure \ref{fig:lengths} shows that the Sweet-Parker width can be as low as 1 cm near the top of the chromosphere, and hence potentially smaller than the ion inertial scale, and so it is possible for Hall effects to be important in some regions of the chromosphere.

The values of reconnection rate in this simulation are comparable to estimated rates from observations of spicules. This fact along with the fact that we see outflows of 3 km/s, comparable to observations of chromospheric jets, provides strong evidence that weakly ionized magnetic reconnection is capable of explaining the ubiquitous observance of transient phenomena in the chromosphere. The plasmoids observed in this simulation, when the current sheet  becomes unstable to the secondary tearing mode (plasmoid instability), may also explain the observations of ``blobs" of plasma within chromospheric jet outflows \citep{2007Sci...318.1591S,2012ApJ...760...28S,2012ApJ...759...33S}. This paper shows that non-equilibrium two fluid (ion + neutral) effects can be important for reconnection on scales larger than those at which Hall and kinetic effects become important in the chromosphere. Hence it is vital that studies of chromospheric reconnection, and the subsequent dynamics and heating this reconnection causes, include ion-neutral collisional physics, as has been done in the model presented in this paper.

\begin{acknowledgments}
\nind{Acknowledgements:}
This work has been supported by the NASA Living With a Star \& Solar and Heliospheric 
Physics programs, the ONR 6.1 Program, and by the NRL-\textit{Hinode} analysis program.
The simulations were performed under a grant of computer time from the DoD HPC program.
\end{acknowledgments}

\providecommand{\noopsort}[1]{}\providecommand{\singleletter}[1]{#1}%
%


\begin{thebibliography}{10}%
\makeatletter
\providecommand \@ifxundefined [1]{%
 \ifx #1\undefined \expandafter \@firstoftwo
 \else \expandafter \@secondoftwo
\fi
}%
\providecommand \@ifnum [1]{%
 \ifnum #1\expandafter \@firstoftwo
 \else \expandafter \@secondoftwo
\fi
}%
\providecommand \enquote [1]{``#1''}%
\providecommand \bibnamefont  [1]{#1}%
\providecommand \bibfnamefont [1]{#1}%
\providecommand \citenamefont [1]{#1}%
\providecommand\href[0]{\@sanitize\@href}%
\providecommand\@href[1]{\endgroup\@@startlink{#1}\endgroup\@@href}%
\providecommand\@@href[1]{#1\@@endlink}%
\providecommand \@sanitize [0]{\begingroup\catcode`\&12\catcode`\#12\relax}%
\@ifxundefined \pdfoutput {\@firstoftwo}{%
 \@ifnum{\z@=\pdfoutput}{\@firstoftwo}{\@secondoftwo}%
}{%
 \providecommand\@@startlink[1]{\leavevmode}%
 \providecommand\@@endlink[0]{}%
}{%
 \providecommand\@@startlink[1]{%
  \leavevmode
  \pdfstartlink
   attr{/Border[0 0 1 ]/H/I/C[0 1 1]}%
   user{/Subtype/Link/A<</Type/Action/S/URI/URI(#1)>>}%
  \relax
 }%
 \providecommand\@@endlink[0]{\pdfendlink}%
}%
\providecommand \url  [0]{\begingroup\@sanitize \@url }%
\providecommand \@url [1]{\endgroup\@href {#1}{\urlprefix}}%
\providecommand \urlprefix [0]{URL }%
\providecommand \Eprint[0]{\href }%
\@ifxundefined \urlstyle {%
  \providecommand \doi [1]{doi:\discretionary{}{}{}#1}%
}{%
  \providecommand \doi [0]{doi:\discretionary{}{}{}\begingroup
  \urlstyle{rm}\Url }%
}%
\providecommand \doibase [0]{http://dx.doi.org/}%
\providecommand \Doi[1]{\href{\doibase#1}}%
\providecommand \selectlanguage [0]{\@gobble}%
\providecommand \bibinfo [0]{\@secondoftwo}%
\providecommand \bibfield [0]{\@secondoftwo}%
\providecommand \translation [1]{[#1]}%
\providecommand \BibitemOpen[0]{}%
\providecommand \bibitemStop [0]{}%
\providecommand \bibitemNoStop [0]{.\EOS\space}%
\providecommand \EOS [0]{\spacefactor3000\relax}%
\providecommand \BibitemShut [1]{\csname bibitem#1\endcsname}%
\bibitem{2009ARA&A..47..291Z}%
  \BibitemOpen
  \bibfield{author}{%
  \bibinfo {author} {\bibfnamefont{E.~G.}\ \bibnamefont{{Zweibel}}}\ and\
  \bibinfo {author} {\bibfnamefont{M.}~\bibnamefont{{Yamada}}},\ }%
  \bibfield{title}{%
  \enquote{\bibinfo {title} {{Magnetic Reconnection in Astrophysical and
  Laboratory Plasmas}},}\ }%
  \bibfield{journal}{%
  \Doi{10.1146/annurev-astro-082708-101726}{\bibinfo {journal} {ArAA}}\ }%
  \textbf{\bibinfo {volume} {47}},\ \bibinfo {pages} {291--332} (\bibinfo
  {month} {Sep.}\ \bibinfo {year} {2009})\BibitemShut{NoStop}%
\bibitem{2011ApJ...731L..18M}%
  \BibitemOpen
  \bibfield{author}{%
  \bibinfo {author} {\bibfnamefont{R.~L.}\ \bibnamefont{{Moore}}}, \bibinfo
  {author} {\bibfnamefont{A.~C.}\ \bibnamefont{{Sterling}}}, \bibinfo {author}
  {\bibfnamefont{J.~W.}\ \bibnamefont{{Cirtain}}},\ and\ \bibinfo {author}
  {\bibfnamefont{D.~A.}\ \bibnamefont{{Falconer}}},\ }%
  \bibfield{title}{%
  \enquote{\bibinfo {title} {{Solar X-ray Jets, Type-II Spicules, Granule-size
  Emerging Bipoles, and the Genesis of the Heliosphere}},}\ }%
  \bibfield{journal}{%
  \Doi{10.1088/2041-8205/731/1/L18}{\bibinfo {journal} {ApJl}}\ }%
  \textbf{\bibinfo {volume} {731}},\ \bibinfo {eid} {L18} (\bibinfo {month}
  {Apr.}\ \bibinfo {year} {2011})\BibitemShut{NoStop}%
\bibitem{2007JGRA..11206215B}%
  \BibitemOpen
  \bibfield{author}{%
  \bibinfo {author} {\bibfnamefont{A.~L.}\ \bibnamefont{{Borg}}}, \bibinfo
  {author} {\bibfnamefont{N.}~\bibnamefont{{{\O}stgaard}}}, \bibinfo {author}
  {\bibfnamefont{A.}~\bibnamefont{{Pedersen}}}, \bibinfo {author}
  {\bibfnamefont{M.}~\bibnamefont{{{\O}ieroset}}}, \bibinfo {author}
  {\bibfnamefont{T.~D.}\ \bibnamefont{{Phan}}}, \bibinfo {author}
  {\bibfnamefont{G.}~\bibnamefont{{Germany}}}, \bibinfo {author}
  {\bibfnamefont{A.}~\bibnamefont{{Aasnes}}}, \bibinfo {author}
  {\bibfnamefont{W.}~\bibnamefont{{Lewis}}}, \bibinfo {author}
  {\bibfnamefont{J.}~\bibnamefont{{Stadsnes}}}, \bibinfo {author}
  {\bibfnamefont{E.~A.}\ \bibnamefont{{Lucek}}}, \bibinfo {author}
  {\bibfnamefont{H.}~\bibnamefont{{R{\`e}me}}},\ and\ \bibinfo {author}
  {\bibfnamefont{C.}~\bibnamefont{{Mouikis}}},\ }%
  \bibfield{title}{%
  \enquote{\bibinfo {title} {{Simultaneous observations of magnetotail
  reconnection and bright X-ray aurora on 2 October 2002}},}\ }%
  \bibfield{journal}{%
  \Doi{10.1029/2006JA011913}{\bibinfo {journal} {Journal of Geophysical
  Research (Space Physics)}}\ }%
  \textbf{\bibinfo {volume} {112}},\ \bibinfo {eid} {A06215} (\bibinfo {month}
  {Jun.}\ \bibinfo {year} {2007})\BibitemShut{NoStop}%
\bibitem{2007JGRA..11201209M}%
  \BibitemOpen
  \bibfield{author}{%
  \bibinfo {author} {\bibfnamefont{S.~E.}\ \bibnamefont{{Milan}}}, \bibinfo
  {author} {\bibfnamefont{G.}~\bibnamefont{{Provan}}},\ and\ \bibinfo {author}
  {\bibfnamefont{B.}~\bibnamefont{{Hubert}}},\ }%
  \bibfield{title}{%
  \enquote{\bibinfo {title} {{Magnetic flux transport in the Dungey cycle: A
  survey of dayside and nightside reconnection rates}},}\ }%
  \bibfield{journal}{%
  \Doi{10.1029/2006JA011642}{\bibinfo {journal} {Journal of Geophysical
  Research (Space Physics)}}\ }%
  \textbf{\bibinfo {volume} {112}},\ \bibinfo {pages} {1209} (\bibinfo {month}
  {Jan.}\ \bibinfo {year} {2007})\BibitemShut{NoStop}%
\bibitem{2005JGRA..11011103E}%
  \BibitemOpen
  \bibfield{author}{%
  \bibinfo {author} {\bibfnamefont{A.~G.}\ \bibnamefont{{Emslie}}}, \bibinfo
  {author} {\bibfnamefont{B.~R.}\ \bibnamefont{{Dennis}}}, \bibinfo {author}
  {\bibfnamefont{G.~D.}\ \bibnamefont{{Holman}}},\ and\ \bibinfo {author}
  {\bibfnamefont{H.~S.}\ \bibnamefont{{Hudson}}},\ }%
  \bibfield{title}{%
  \enquote{\bibinfo {title} {{Refinements to flare energy estimates: A followup
  to ``Energy partition in two solar flare/CME events'' by A. G. Emslie et
  al.}}.}\ }%
  \bibfield{journal}{%
  \Doi{10.1029/2005JA011305}{\bibinfo {journal} {Journal of Geophysical
  Research (Space Physics)}}\ }%
  \textbf{\bibinfo {volume} {110}},\ \bibinfo {eid} {A11103} (\bibinfo {month}
  {Nov.}\ \bibinfo {year} {2005})\BibitemShut{NoStop}%
\bibitem{2005ApJ...628L..77T}%
  \BibitemOpen
  \bibfield{author}{%
  \bibinfo {author} {\bibfnamefont{S.}~\bibnamefont{{Tanuma}}}\ and\ \bibinfo
  {author} {\bibfnamefont{K.}~\bibnamefont{{Shibata}}},\ }%
  \bibfield{title}{%
  \enquote{\bibinfo {title} {{Internal Shocks in the Magnetic Reconnection Jet
  in Solar Flares: Multiple Fast Shocks Created by the Secondary Tearing
  Instability}},}\ }%
  \bibfield{journal}{%
  \Doi{10.1086/432418}{\bibinfo {journal} {ApJl}}\ }%
  \textbf{\bibinfo {volume} {628}},\ \bibinfo {pages} {L77--L80} (\bibinfo
  {month} {Jul.}\ \bibinfo {year} {2005}),\
  \Eprint{http://arxiv.org/abs/arXiv:astro-ph/0503005}{arXiv:astro-ph/0503005}%
\BibitemShut{NoStop}%
\bibitem{1957JGR....62..509P}%
  \BibitemOpen
  \bibfield{author}{%
  \bibinfo {author} {\bibfnamefont{E.~N.}\ \bibnamefont{{Parker}}},\ }%
  \bibfield{title}{%
  \enquote{\bibinfo {title} {{Sweet's Mechanism for Merging Magnetic Fields in
  Conducting Fluids}},}\ }%
  \bibfield{journal}{%
  \Doi{10.1029/JZ062i004p00509}{\bibinfo {journal} {JGR}}\ }%
  \textbf{\bibinfo {volume} {62}},\ \bibinfo {pages} {509--520} (\bibinfo
  {month} {Dec.}\ \bibinfo {year} {1957})\BibitemShut{NoStop}%
\bibitem{Sweet58}%
  \BibitemOpen
  \bibfield{author}{%
  \bibinfo {author} {\bibfnamefont{P.~A.}\ \bibnamefont{Sweet}},\ }%
  \enquote{\bibinfo {title} {Electromagnetic phenomena in cosmical physics},}\
  \ (\bibinfo {publisher} {Cambridge U.P., New York},\ \bibinfo {year} {1958})\
  p.\ \bibinfo {pages} {123}\BibitemShut{NoStop}%
\bibitem{2007Sci...318.1591S}%
  \BibitemOpen
  \bibfield{author}{%
  \bibinfo {author} {\bibfnamefont{K.}~\bibnamefont{{Shibata}}}, \bibinfo
  {author} {\bibfnamefont{T.}~\bibnamefont{{Nakamura}}}, \bibinfo {author}
  {\bibfnamefont{T.}~\bibnamefont{{Matsumoto}}}, \bibinfo {author}
  {\bibfnamefont{K.}~\bibnamefont{{Otsuji}}}, \bibinfo {author}
  {\bibfnamefont{T.~J.}\ \bibnamefont{{Okamoto}}}, \bibinfo {author}
  {\bibfnamefont{N.}~\bibnamefont{{Nishizuka}}}, \bibinfo {author}
  {\bibfnamefont{T.}~\bibnamefont{{Kawate}}}, \bibinfo {author}
  {\bibfnamefont{H.}~\bibnamefont{{Watanabe}}}, \bibinfo {author}
  {\bibfnamefont{S.}~\bibnamefont{{Nagata}}}, \bibinfo {author}
  {\bibfnamefont{S.}~\bibnamefont{{UeNo}}}, \bibinfo {author}
  {\bibfnamefont{R.}~\bibnamefont{{Kitai}}}, \bibinfo {author}
  {\bibfnamefont{S.}~\bibnamefont{{Nozawa}}}, \bibinfo {author}
  {\bibfnamefont{S.}~\bibnamefont{{Tsuneta}}}, \bibinfo {author}
  {\bibfnamefont{Y.}~\bibnamefont{{Suematsu}}}, \bibinfo {author}
  {\bibfnamefont{K.}~\bibnamefont{{Ichimoto}}}, \bibinfo {author}
  {\bibfnamefont{T.}~\bibnamefont{{Shimizu}}}, \bibinfo {author}
  {\bibfnamefont{Y.}~\bibnamefont{{Katsukawa}}}, \bibinfo {author}
  {\bibfnamefont{T.~D.}\ \bibnamefont{{Tarbell}}}, \bibinfo {author}
  {\bibfnamefont{T.~E.}\ \bibnamefont{{Berger}}}, \bibinfo {author}
  {\bibfnamefont{B.~W.}\ \bibnamefont{{Lites}}}, \bibinfo {author}
  {\bibfnamefont{R.~A.}\ \bibnamefont{{Shine}}},\ and\ \bibinfo {author}
  {\bibfnamefont{A.~M.}\ \bibnamefont{{Title}}},\ }%
  \bibfield{title}{%
  \enquote{\bibinfo {title} {{Chromospheric Anemone Jets as Evidence of
  Ubiquitous Reconnection}},}\ }%
  \bibfield{journal}{%
  \Doi{10.1126/science.1146708}{\bibinfo {journal} {Science}}\ }%
  \textbf{\bibinfo {volume} {318}},\ \bibinfo {pages} {1591--} (\bibinfo
  {month} {Dec.}\ \bibinfo {year} {2007}),\
  \Eprint{http://arxiv.org/abs/0810.3974}{arXiv:0810.3974}\BibitemShut{NoStop}%
\bibitem{2012ApJ...760...28S}%
  \BibitemOpen
  \bibfield{author}{%
  \bibinfo {author} {\bibfnamefont{K.~A.~P.}\ \bibnamefont{{Singh}}}, \bibinfo
  {author} {\bibfnamefont{H.}~\bibnamefont{{Isobe}}}, \bibinfo {author}
  {\bibfnamefont{K.}~\bibnamefont{{Nishida}}},\ and\ \bibinfo {author}
  {\bibfnamefont{K.}~\bibnamefont{{Shibata}}},\ }%
  \bibfield{title}{%
  \enquote{\bibinfo {title} {{Systematic Motion of Fine-scale Jets and
  Successive Reconnection in Solar Chromospheric Anemone Jet Observed with the
  Solar Optical Telescope/Hinode}},}\ }%
  \bibfield{journal}{%
  \Doi{10.1088/0004-637X/760/1/28}{\bibinfo {journal} {ApJ}}\ }%
  \textbf{\bibinfo {volume} {760}},\ \bibinfo {eid} {28} (\bibinfo {month}
  {Nov.}\ \bibinfo {year} {2012})\BibitemShut{NoStop}%
\bibitem{2012ApJ...759...33S}%
  \BibitemOpen
  \bibfield{author}{%
  \bibinfo {author} {\bibfnamefont{K.~A.~P.}\ \bibnamefont{{Singh}}}, \bibinfo
  {author} {\bibfnamefont{H.}~\bibnamefont{{Isobe}}}, \bibinfo {author}
  {\bibfnamefont{N.}~\bibnamefont{{Nishizuka}}}, \bibinfo {author}
  {\bibfnamefont{K.}~\bibnamefont{{Nishida}}},\ and\ \bibinfo {author}
  {\bibfnamefont{K.}~\bibnamefont{{Shibata}}},\ }%
  \bibfield{title}{%
  \enquote{\bibinfo {title} {{Multiple Plasma Ejections and Intermittent Nature
  of Magnetic Reconnection in Solar Chromospheric Anemone Jets}},}\ }%
  \bibfield{journal}{%
  \Doi{10.1088/0004-637X/759/1/33}{\bibinfo {journal} {ApJ}}\ }%
  \textbf{\bibinfo {volume} {759}},\ \bibinfo {eid} {33} (\bibinfo {month}
  {Nov.}\ \bibinfo {year} {2012})\BibitemShut{NoStop}%
\bibitem{2000SoPh..196...79S}%
  \BibitemOpen
  \bibfield{author}{%
  \bibinfo {author} {\bibfnamefont{A.~C.}\ \bibnamefont{{Sterling}}},\ }%
  \bibfield{title}{%
  \enquote{\bibinfo {title} {{Solar Spicules: A Review of Recent Models and
  Targets for Future Observations - (Invited Review)}},}\ }%
  \bibfield{journal}{%
  \bibinfo {journal} {Solar Physics}\ }%
  \textbf{\bibinfo {volume} {196}},\ \bibinfo {pages} {79--111} (\bibinfo
  {month} {Sep.}\ \bibinfo {year} {2000})\BibitemShut{NoStop}%
\bibitem{Leake2012}%
  \BibitemOpen
  \bibfield{author}{%
  \bibinfo {author} {\bibfnamefont{J.~E.}\ \bibnamefont{{Leake}}}, \bibinfo
  {author} {\bibfnamefont{V.~S.}\ \bibnamefont{{Lukin}}}, \bibinfo {author}
  {\bibfnamefont{M.~G.}\ \bibnamefont{{Linton}}},\ and\ \bibinfo {author}
  {\bibfnamefont{E.~T.}\ \bibnamefont{{Meier}}},\ }%
  \bibfield{title}{%
  \enquote{\bibinfo {title} {{Multi-fluid Simulations of Chromospheric Magnetic
  Reconnection in a Weakly Ionized Reacting Plasma}},}\ }%
  \bibfield{journal}{%
  \Doi{10.1088/0004-637X/760/2/109}{\bibinfo {journal} {ApJ}}\ }%
  \textbf{\bibinfo {volume} {760}},\ \bibinfo {eid} {109} (\bibinfo {month}
  {Dec.}\ \bibinfo {year} {2012}),\
  \Eprint{http://arxiv.org/abs/1210.1807}{arXiv:1210.1807
  [physics.plasm-ph]}\BibitemShut{NoStop}%
\bibitem{Meier11}%
  \BibitemOpen
  \bibfield{author}{%
  \bibinfo {author} {\bibfnamefont{E.~T.}\ \bibnamefont{Meier}},\ }%
  \emph{\bibinfo {title} {Modeling Plasmas with Strong Anisotropy, Neutral
  Fluid Effects, and Open Boundaries}},\ Ph.D. thesis,\ \bibinfo {school}
  {University of Washington} (\bibinfo {year} {2011})\BibitemShut{NoStop}%
\bibitem{1990STIN...9113238B}%
  \BibitemOpen
  \bibfield{author}{%
  \bibinfo {author} {\bibfnamefont{C.~F.}\ \bibnamefont{{Barnett}}}, \bibinfo
  {author} {\bibfnamefont{H.~T.}\ \bibnamefont{{Hunter}}}, \bibinfo {author}
  {\bibfnamefont{M.~I.}\ \bibnamefont{{Fitzpatrick}}}, \bibinfo {author}
  {\bibfnamefont{I.}~\bibnamefont{{Alvarez}}}, \bibinfo {author}
  {\bibfnamefont{C.}~\bibnamefont{{Cisneros}}},\ and\ \bibinfo {author}
  {\bibfnamefont{R.~A.}\ \bibnamefont{{Phaneuf}}},\ }%
  \bibfield{title}{%
  \enquote{\bibinfo {title} {{Atomic data for fusion. Volume 1: Collisions of
  H, H2, He and Li atoms and ions with atoms and molecules}},}\ }%
  \bibfield{journal}{%
  \bibinfo {journal} {NASA STI/Recon Technical Report N}\ }%
  \textbf{\bibinfo {volume} {91}},\ \bibinfo {pages} {13238} (\bibinfo {month}
  {Jul.}\ \bibinfo {year} {1990})\BibitemShut{NoStop}%
\bibitem{1983ApJ...264..485D}%
  \BibitemOpen
  \bibfield{author}{%
  \bibinfo {author} {\bibfnamefont{B.~T.}\ \bibnamefont{{Draine}}}, \bibinfo
  {author} {\bibfnamefont{W.~G.}\ \bibnamefont{{Roberge}}},\ and\ \bibinfo
  {author} {\bibfnamefont{A.}~\bibnamefont{{Dalgarno}}},\ }%
  \bibfield{title}{%
  \enquote{\bibinfo {title} {{Magnetohydrodynamic shock waves in molecular
  clouds}},}\ }%
  \bibfield{journal}{%
  \Doi{10.1086/160617}{\bibinfo {journal} {ApJ}}\ }%
  \textbf{\bibinfo {volume} {264}},\ \bibinfo {pages} {485--507} (\bibinfo
  {month} {Jan.}\ \bibinfo {year} {1983})\BibitemShut{NoStop}%
\bibitem{1981ApJS...45..635V}%
  \BibitemOpen
  \bibfield{author}{%
  \bibinfo {author} {\bibfnamefont{J.~E.}\ \bibnamefont{{Vernazza}}}, \bibinfo
  {author} {\bibfnamefont{E.~H.}\ \bibnamefont{{Avrett}}},\ and\ \bibinfo
  {author} {\bibfnamefont{R.}~\bibnamefont{{Loeser}}},\ }%
  \bibfield{title}{%
  \enquote{\bibinfo {title} {{Structure of the solar chromosphere. III - Models
  of the EUV brightness components of the quiet-sun}},}\ }%
  \bibfield{journal}{%
  \Doi{10.1086/190731}{\bibinfo {journal} {ApJs}}\ }%
  \textbf{\bibinfo {volume} {45}},\ \bibinfo {pages} {635--725} (\bibinfo
  {month} {Apr.}\ \bibinfo {year} {1981})\BibitemShut{NoStop}%
\bibitem{1999ApJ...511..193V}%
  \BibitemOpen
  \bibfield{author}{%
  \bibinfo {author} {\bibfnamefont{E.~T.}\ \bibnamefont{{Vishniac}}}\ and\
  \bibinfo {author} {\bibfnamefont{A.}~\bibnamefont{{Lazarian}}},\ }%
  \bibfield{title}{%
  \enquote{\bibinfo {title} {{Reconnection in the Interstellar Medium}},}\ }%
  \bibfield{journal}{%
  \Doi{10.1086/306643}{\bibinfo {journal} {ApJ}}\ }%
  \textbf{\bibinfo {volume} {511}},\ \bibinfo {pages} {193--203} (\bibinfo
  {month} {Jan.}\ \bibinfo {year} {1999})\BibitemShut{NoStop}%
\bibitem{2003ApJ...583..229H}%
  \BibitemOpen
  \bibfield{author}{%
  \bibinfo {author} {\bibfnamefont{F.}~\bibnamefont{{Heitsch}}}\ and\ \bibinfo
  {author} {\bibfnamefont{E.~G.}\ \bibnamefont{{Zweibel}}},\ }%
  \bibfield{title}{%
  \enquote{\bibinfo {title} {{Fast Reconnection in a Two-Stage Process}},}\ }%
  \bibfield{journal}{%
  \Doi{10.1086/345082}{\bibinfo {journal} {ApJ}}\ }%
  \textbf{\bibinfo {volume} {583}},\ \bibinfo {pages} {229--244} (\bibinfo
  {month} {Jan.}\ \bibinfo {year} {2003}),\
  \Eprint{http://arxiv.org/abs/arXiv:astro-ph/0205103}{arXiv:astro-ph/0205103}%
\BibitemShut{NoStop}%
\bibitem{Loureiro07}%
  \BibitemOpen
  \bibfield{author}{%
  \bibinfo {author} {\bibfnamefont{N.~F.}\ \bibnamefont{Loureiro}}, \bibinfo
  {author} {\bibfnamefont{A.~A.}\ \bibnamefont{Schekochihin}},\ and\ \bibinfo
  {author} {\bibfnamefont{S.~C.}\ \bibnamefont{Cowley}},\ }%
  \bibfield{title}{%
  \enquote{\bibinfo {title} {Instability of current sheets and formation of
  plasmoid chains},}\ }%
  \bibfield{journal}{%
  \bibinfo {journal} {Phys. Plasmas}\ }%
  \textbf{\bibinfo {volume} {14}},\ \bibinfo {pages} {100703} (\bibinfo {year}
  {2007})\BibitemShut{NoStop}%
\bibitem{1978JETPL..28..177B}%
  \BibitemOpen
  \bibfield{author}{%
  \bibinfo {author} {\bibfnamefont{S.~V.}\ \bibnamefont{{Bulanov}}}, \bibinfo
  {author} {\bibfnamefont{S.~I.}\ \bibnamefont{{Syrovatski{\v i}}}},\ and\
  \bibinfo {author} {\bibfnamefont{J.}~\bibnamefont{{Sakai}}},\ }%
  \bibfield{title}{%
  \enquote{\bibinfo {title} {{Stabilizing influence of plasma flow on
  dissipative tearing instability}},}\ }%
  \bibfield{journal}{%
  \bibinfo {journal} {Soviet Journal of Experimental and Theoretical Physics
  Letters}\ }%
  \textbf{\bibinfo {volume} {28}},\ \bibinfo {pages} {177} (\bibinfo {month}
  {Aug.}\ \bibinfo {year} {1978})\BibitemShut{NoStop}%
\bibitem{2006ApJ...639..441F}%
  \BibitemOpen
  \bibfield{author}{%
  \bibinfo {author} {\bibfnamefont{J.~M.}\ \bibnamefont{{Fontenla}}}, \bibinfo
  {author} {\bibfnamefont{E.}~\bibnamefont{{Avrett}}}, \bibinfo {author}
  {\bibfnamefont{G.}~\bibnamefont{{Thuillier}}},\ and\ \bibinfo {author}
  {\bibfnamefont{J.}~\bibnamefont{{Harder}}},\ }%
  \bibfield{title}{%
  \enquote{\bibinfo {title} {{Semiempirical Models of the Solar Atmosphere. I.
  The Quiet- and Active Sun Photosphere at Moderate Resolution}},}\ }%
  \bibfield{journal}{%
  \Doi{10.1086/499345}{\bibinfo {journal} {ApJ}}\ }%
  \textbf{\bibinfo {volume} {639}},\ \bibinfo {pages} {441--458} (\bibinfo
  {month} {Mar.}\ \bibinfo {year} {2006})\BibitemShut{NoStop}%
\bibitem{2012PhPl...19g2902N}%
  \BibitemOpen
  \bibfield{author}{%
  \bibinfo {author} {\bibfnamefont{L.}~\bibnamefont{{Ni}}}, \bibinfo {author}
  {\bibfnamefont{U.}~\bibnamefont{{Ziegler}}}, \bibinfo {author}
  {\bibfnamefont{Y.-M.}\ \bibnamefont{{Huang}}}, \bibinfo {author}
  {\bibfnamefont{J.}~\bibnamefont{{Lin}}},\ and\ \bibinfo {author}
  {\bibfnamefont{Z.}~\bibnamefont{{Mei}}},\ }%
  \bibfield{title}{%
  \enquote{\bibinfo {title} {{Effects of plasma {$\beta$} on the plasmoid
  instability}},}\ }%
  \bibfield{journal}{%
  \Doi{10.1063/1.4736993}{\bibinfo {journal} {Physics of Plasmas}}\ }%
  \textbf{\bibinfo {volume} {19}},\ \bibinfo {pages} {072902} (\bibinfo {month}
  {Jul.}\ \bibinfo {year} {2012})\BibitemShut{NoStop}%
\bibitem{Birn01}%
  \BibitemOpen
  \bibfield{author}{%
  \bibinfo {author} {\bibfnamefont{J.}~\bibnamefont{Birn}}, \bibinfo {author}
  {\bibfnamefont{J.~F.}\ \bibnamefont{Drake}}, \bibinfo {author}
  {\bibfnamefont{M.~A.}\ \bibnamefont{Shay}}, \bibinfo {author}
  {\bibfnamefont{B.~N.}\ \bibnamefont{Rogers}}, \bibinfo {author}
  {\bibfnamefont{R.~E.}\ \bibnamefont{Denton}}, \bibinfo {author}
  {\bibfnamefont{M.}~\bibnamefont{Hesse}}, \bibinfo {author}
  {\bibfnamefont{M.}~\bibnamefont{Kuznetsova}}, \bibinfo {author}
  {\bibfnamefont{Z.~W.}\ \bibnamefont{Ma}}, \bibinfo {author}
  {\bibfnamefont{A.}~\bibnamefont{Bhattacharjee}}, \bibinfo {author}
  {\bibfnamefont{A.}~\bibnamefont{Otto}},\ and\ \bibinfo {author}
  {\bibfnamefont{P.~L.}\ \bibnamefont{Pritchett}},\ }%
  \bibfield{title}{%
  \enquote{\bibinfo {title} {{Geospace Environment Modeling (GEM)} magnetic
  reconnection challenge: resistive tearing, anisotropic pressure and {Hall}
  effects},}\ }%
  \bibfield{journal}{%
  \bibinfo {journal} {J. Geophys. Res.}\ }%
  \textbf{\bibinfo {volume} {106}},\ \bibinfo {pages} {3715} (\bibinfo {year}
  {2001})\BibitemShut{NoStop}%
\bibitem{Huang10}%
  \BibitemOpen
  \bibfield{author}{%
  \bibinfo {author} {\bibfnamefont{Y.-M.}\ \bibnamefont{Huang}}\ and\ \bibinfo
  {author} {\bibfnamefont{A.}~\bibnamefont{Bhattacharjee}},\ }%
  \bibfield{title}{%
  \enquote{\bibinfo {title} {Scaling laws of resistive magnetohydrodynamic
  reconnection in the high-lundquist-number, plasmoid-unstable regime},}\ }%
  \bibfield{journal}{%
  \bibinfo {journal} {Phys. Plasmas}\ }%
  \textbf{\bibinfo {volume} {17}},\ \bibinfo {pages} {062104} (\bibinfo {year}
  {2010})\BibitemShut{NoStop}%
\end{thebibliography}

\end{document}